\documentclass[prb,twocolumn,preprintnumbers,amsmath,amssymb,showpacs,superscriptaddress,floatfix,dvips]{revtex4}

\usepackage[dvipdfmx]{graphicx}
\usepackage{dcolumn}
\usepackage{bm}
\usepackage{ulem}
\usepackage[dvips]{color}

\begin{document}

%
\title{Metal-insulator transition in the 2D Hubbard model:\\ dual fermion approach with the Lanczos exact diagonalization}
\author{Arata Tanaka}%
\affiliation{Department of Quantum Matter, ADSM, Hiroshima University, Higashi-hiroshima 739-8530, Japan}
\date{\today}
\begin{abstract}
In this study the metal-insulator transition in the square-lattice Hubbard model at half-filling is revisited in relation to the DOS and spectral functions by means of the ladder dual fermion approximation (LDFA).  For this purpose, a new expression of the two-body Green's function in the form of resolvents is proposed, which provides tractable and efficient means to calculate the local vertex function with the Lanczos exact diagonalization (ED) method. This makes it possible to use the Lanczos ED method as a solver of the effective impurity Anderson model for LDFA, opening up the way to access low temperatures for these perturbative extensions of the dynamical mean field theory and to obtain accurate DOS and spectral functions on the real frequency axis by a new variant of the maximum entropy method.  It is found that for $U\le 3.5t$, as temperature decreases, the pseudogap formation due to antiferromagnetic correlations in the quasiparticle peak of the spectral function occurs at the X point $[\bm{k}=(\pi,0)]$, spreads through the Fermi surface and ends at the M$_2$ point $[\bm{k}=(\pi/2,\pi/2)]$. The almost simultaneous creation of the pseudogap and the loss of the Fermi liquid feature is consistent with that expected in the Slater regime.  Although the pseudogap still appears in the quasi-particle-like single peak for $U\ge 4.0t$, the Fermi-liquid feature is partially lost on the Fermi surface already at higher temperatures as expected in the Mott-Heisenberg regime, in which local spins are preformed at high temperatures. A sharp crossover from a pseudogap phase to a Mott insulator at finite $U^{*} \approx 4.7t$ is found to occur below the temperature of the pseudogap formation similar to a previous study with the non-linear $\sigma$ model approach.
\end{abstract}
\pacs{}
\maketitle
\section{Introduction}
Nowadays the dynamical mean field theory (DMFT)\cite{AGeorges1996} is one of the most powerful methods to describe electronic properties in strongly correlated electron systems. Although DMFT is exact in infinite dimension\cite{WMetzner1989}, spatial correlation effects, which are absence in DMFT similar to the ordinary mean-field theory, play crucial roles in finite dimensions in some fascinating cases, such as criticality due to thermal or quantum fluctuations, unconventional superconductors, and metal-insulator transitions (MIT) in low dimensions. In recent years, to include spatial correlations, numerous extensions of DMFT have been developed. In DMFT, the problem of electrons on the lattice is mapped to the effective impurity Anderson model (IAM), where the interaction is explicitly considered on one of the sites (the impurity site) and rest of the sites are replaced by an effective medium. In the cluster extensions of DMFT such as the cellular DMFT (CDMFT) or the dynamical cluster approximation (DCA)\cite{TMaier2005}, the impurity site is replaced by a cluster and spatial correlations within the cluster are considered.  

Other attempts to include spatial correlations are perturbative extensions of DMFT\cite{GRohringer2018} and among of them, there is a class of methods in which the local vertex functions instead of the bare interaction are used as the diagrammatic elements of the perturbation. Those in this category are, for example, the pioneering work of Kusunose\cite{HKusunose2006}, the dynamical vertex approximation (D$\Gamma$A)\cite{AToschi2007}, the dual fermion approximation (DFA)\cite{ANRubtsov2008,ANRubtsov2009}, the dual boson approximation (DBA)\cite{ANRubtsov2012}, the one-particle irreducible approach (1PI)\cite{GRohringer2013}, the DMFT to functional renormalization group approach (DMFT$^2$RG)\cite{CTaranto2014}, and the triply irreducible local expansion (TRILEX)\cite{TAyral2015}.

To solve the effective IAM for DMFT, various numerical methods have been developed. In the perturbative extensions of DMFT mentioned above, the two-body Green's function of the effective IAM is further required to obtain the local four-point vertex function. Efficient schemes to calculate the two-body Green's function have already been developed and applied to DFA and D$\Gamma$A with the continuous-time quantum Monte Carlo (QMC)\cite{GLi2008,HHafermann2012,HHafermann2014,HShinaoka2018} and the ordinary exact diagonalization (ED)\cite{AToschi2007,HHafermann2009b,HHafermann2010,GRohringer2012,AValli2015} methods. The QMC methods have difficulty in accuracy particularly in low temperatures because of the statical errors. On the other hand, for the ordinary ED technique, the method so far proposed is that with the Lehmann representation and as will be discussed later, it has problem in efficiency and the limitation of the number of the many-body basis functions. 

One of the purposes of this paper is to present a new formula for the two-body Green's function. This renders efficient and accurate means to calculate the local vertex function and makes it possible to use the Lanczos ED technique\cite{AGeorges1996,ZBai2000} as a solver of the effective IAM required for DFA and similar perturbative extensions of DMFT. 

The MIT in the two-dimensional (2D) Hubbard model at half filling still remains to be a subject of debate even after decades of extensive studies\cite{JRSchrieffer1989,MVekic1993,YMVilk1996,YMVilk1997,PWAnderson1997,SMoukouri2000,FMancini2000,AAvella2001,KBorejsza2003,KBorejsza2004}. 
Although it is well established that the ground state of the half-filled 2D Hubbard model on a square lattice has long-range antiferromagnetic (AFM) order\cite{JEHirsch1985,SRWhite1989}, the difficulty mainly arises from the fact that long-range AFM order cannot be stable at finite temperatures in two dimension because of the Mermin-Wagner theorem\cite{NDMermin1966}. The MIT has been discussed in relation to the Slater and Mott-Heisenberg mechanisms. In the Slater regime, the gap formation is essentially that of one-body picture, i.e., the Brillouin zone folding caused by the AFM ordering. Hence, the spin and charge degrees of freedom are entwined and both the spin and charge excitations have the same energy scale. In contrast, in the Mott-Heisenberg regime, localized spins preformed at high temperatures and AFM ordering occurs through the exchange coupling between these local spins.  Since this picture is essentially based on many-body theories, the energy scale of the charge excitation $\approx U$ and the spin excitation $\approx 4t^2/U$ are generally different. In a study with the non-linear $\sigma$ model approach\cite{KBorejsza2003,KBorejsza2004}, the pseudogap is formed at low temperatures. While a clear insulating gap opens in the Mott-Heisenberg regime, the DOS at the Fermi level remains finite for $T>0$ in the Slater regime and thus the MIT point $U_{\rm c}\approx 4.2t$ is expected to be positioned at the boundary between these two regimes. On the other hand, Anderson has proposed that whole low temperature physics of the 2D Hubbard model is mapped onto the 2D Heisenberg model and a Mott gap opens for all $U>0$\cite{PWAnderson1997}. 

DMFT predicts the first-order Mott MIT at finite temperatures with a second-order critical endpoint $U_{\rm c}\approx 10t$ at $T=0$ when the paramagnetic (PM) state is assumed\cite{AGeorges1996}, which is essentially the same to the MIT in infinite dimension. CDMFT\cite{HPark2008,LFratino2017}, the variational cluster approximation (VCA)\cite{TSchaefer2015} and the second-order DFA\cite{HHafermann2010,EGCPvanLoon2018b}, which are only capable for short-range spatial correlations within the cluster, also find the first-order MIT at finite temperatures similar to DMFT but with substantially smaller critical values $U_{\rm c}\approx 6t$. In these theories, however, the AFM insulating state have finite N{\'e}el temperatures and the region where the first-order MIT line presence in the $U$-$T$ phase diagram is replaced by the AFM insulating phase when the solutions are not constrained to the PM state\cite{LFratino2017}. 

In the studies by means of D$\Gamma$A and extrapolated lattice QMC\cite{TSchaefer2015,TSchaefer2016}, the ladder dual fermion approximation (LDFA)\cite{GRohringer2018}, and the two-particle self-consistent approximation (TPSC)\cite{YMVilk1996,YMVilk1997}, which incorporate the effects of long-range correlations and fulfill the Mermin-Wagner theorem, the MIT occurs at much smaller $U$ compared to the CDMFT results. In the QMC calculations of finite-size clusters\cite{DRost2012} has also found the pseudogap at least $U\ge 2.0t$. Sch{\"a}fer {\it et.~al.} suggest in their combined study of D$\Gamma$A and lattice QMC\cite{TSchaefer2015} that $U_{\rm c}=0$ for $T\to 0$ and thus no MIT occurs at any $U>0$ similar to the  1D Hubbard model\cite{EHLieb1968}. 

However, the formation of the pseudogap does not necessarily indicate insulating behavior in low temperatures and examination of subtle changes in states inside the gap as a function of temperature is required to verify whether $U_{\rm c}$ stays finite or $U_{\rm c}=0$\cite{KBorejsza2004}. For this reason, it is essential to obtain precise information of the DOS and spectral function in the vicinity of the Fermi level to understand the MIT in the 2D Hubbard model. Although there are already several LDFA works\cite{HHafermann2009,JOtsuki2014,EGCPvanLoon2018,TRibic2018} and that with the diagrammatic Mote Carlo approach\cite{SIskakov2016,JGukelberger2017} on the 2D Hubbard model at half filling, detailed investigation on the DOS and spectral function with LDFA at low temperatures is still lacking.

In addition to a new formula for the two-body Green's function, the other purpose of this paper is to investigate the DOS and spectral function of the square-lattice Hubbard model at half filling by means of LDFA to elucidate the behavior and origin of the MIT. In particular, utilizing the newly developed Lanczos ED scheme to calculate the two-body Green's function as the solver of the effective IAM for LDFA, it is possible to access large $U$ and low temperature region of the $U$-$T$ phase diagram where previous studies still have not reached and obtain results with unprecedented accuracy. It is found that a sharp crossover from a pseudogap phase to a Mott insulator around $U^{*}\approx 4.7t$ occurs below the temperature of the pseudogap formation. 

The rest of this paper is structured as follows. In Sec.~\ref{UT}, a short explanation on the $U$-$T$ phase diagram of the 2D Hubbard model obtained in this study is given. In Sec.~\ref{Expression}, the new formula of two-body Green's function is presented. Section~\ref{LanczosED} describes how to calculate the two-body Green's function approximately with the Lanczos algorithm with the new formula.  In Sec.~\ref{LDFA}, a brief overview of LDFA is given and some technical points specific to the Lanczos ED method are presented. In Sec.~\ref{MEM}, a detailed description of the maximum entropy method used in this study is given. In Sec.~\ref{2DHubbard}, results of LDFA calculations of the 2D Hubbard model are presented. The paper is closed with a discussion in Sec.~\ref{Discussions} and a brief summary in  Sec.~\ref{Conclusions}. Derivations of the new formula in Sec.~\ref{Expression} and the update formula of the hybridization function  in Sec.~\ref{2DHubbard} are deferred to Appendices A and B. The convergence of the vertex function of IAM with the Lanczos ED method is discussed in Appendix C and the DOS of the 2D Hubbard model inferred by the present and standard  maximum entropy methods are compared in Appendix D.

\section{$U$-$T$ phase diagram of the 2D Hubbard model\label{UT}}
\begin{figure}
\includegraphics[width=8cm]{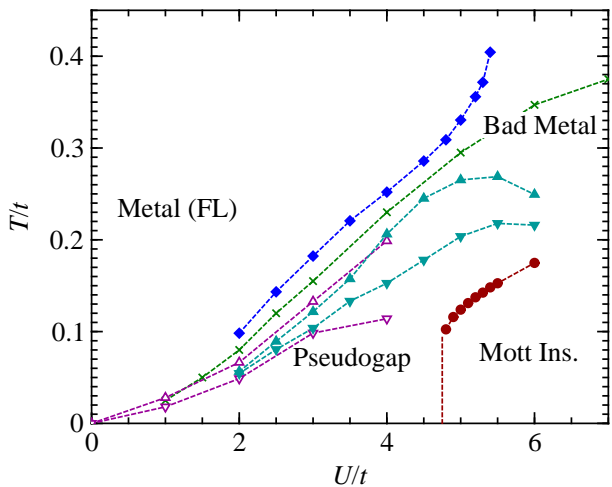}
\caption{\label{UTphase}$U$-$T$ phase diagram for the half-filled Hubbard model on a square lattice obtained with LDFA. Temperatures where the double occupancy $D$ has the local maximum (closed diamonds),  the pseudogap formation occur at the X (closed tip-up triangles) and M$_2$ (closed tip-down triangles) points in the spectral function $A_{\bm{k}}(\omega)$ and the DOS at the Fermi level is $\rho(\omega=0)=10^{-7}$ for $U\ge 4.8t$ (closed circles) are shown. The vertical line at $U=4.7t$ is the crossover line from the pseudogap phase to the Mott insulator. For references, the DMFT N{\'e}el temperature $T^{\rm DMFT}_{\rm N}$ (crosses) in Ref.\,\onlinecite{JOtsuki2014} and the pseudogap formation temperatures at the X (open tip-up triangles) and M$_2$ (open tip-down triangles) points  in the combined study of D$\Gamma$A and lattice QMC\cite{TSchaefer2016} are also presented.}
\end{figure}
Before embarking on rather lengthy explanation of the present computational scheme, here, we give a succinct account on the $U$-$T$ phase diagram of the 2D Hubbard model obtained in this study.  Figure \ref{UTphase} shows $U$-$T$ phase diagram for the half-filled Hubbard model on a square lattice obtained with LDFA. As will be discussed in Sec.~\ref{ssPG}, 
a metallic to pseudogap phase crossover is found to occur with decreasing temperature: the pseudogap is first formed in the quasiparticle peak of the spectral function at the X point $\bm{k}=(\pi,0)$  (shown by the closed tip-up triangles) and the formation spreads through the Fermi surface and ends at the M$_2$ point $\bm{k}=(\pi/2,\pi/2)$ (closed tip-down triangles). The results are consistent with the previous D$\Gamma$A and lattice QMC study\cite{TSchaefer2015} (also indicated in Fig.~\ref{UTphase}). 

However, the characteristic of the pseudogap formation changes depending on the size of $U$.
For $U \le 3.5t$, the pseudogap formation and the loss of the Fermi-liquid feature occur simultaneously and below the temperature of the pseudogap formation at the M$_2$ point, the Fermi surface is totally lost and the system enters the pseudogap phase. These results for $U \le 3.5t$ are consistent with those expected in the Slater regime. 

On the other hand, for larger $U$, although the formation of the pseudogap still occurs in the quasiparticle-like single peak at the Fermi level accompanied by prominent shoulder structures at $\omega\approx\pm U/2$, the Fermi-liquid feature is partially lost around the X point already at higher temperatures above the DMFT N{\'e}el temperature $T^{\rm DMFT}_{\rm N}$ for $U=4.0t$ and totally lost for $U\ge 5.5t$. These results for $U \ge 4.0t$ is consistent with those expected in the Mott-Heisenberg regime, in which local spins are preformed above the temperature where AFM correlations start to develop.

As will be discussed in Sec.~\ref{ssGap}, a sharp crossover from the pseudogap phase to the Mott insulator around $U^{*} \approx 4.7t$ is found to occur below the temperature of the pseudogap formation. For $U < U^{*}$, the DOS at the Fermi level is reduced with decreasing temperature but persists even at low temperatures. In contrast, for $U > U^{*}$, the reduction is much rapid and clear gap opening occurs at certain temperature. The value of $U^{*} \approx 4.7t$ coincides with the boundary between the Slater and Mott-Heisenberg regimes defined by the inflection point of the double occupancy $D$ curve as a function of $U$ as will be discussed in Sec.~\ref{ssD}. These low-energy behavior of DOS in the vicinity of the Fermi level is consistent with the previous study with the non-linear $\sigma$ model approach\cite{KBorejsza2004}.

\section{Expression of two-body green's function in the form of resolvents\label{Expression}}
As an expression for two-body Green's function, the Lehmann representation has been used for the ordinary ED technique\cite{AToschi2007,HHafermann2009b,HHafermann2010}. However, the structure of the formula is not suitable for the Lanczos exact diagonalization method. The expression has terms like
\begin{align}
&\sum_{lmnk}
\frac{\left(e^{-\beta E_l}/Z\right)\langle k|c_2|l\rangle\langle l|c_1|m\rangle\langle m|c^\dagger_3|n\rangle\langle n|c^\dagger_4|k\rangle }{(E_k-E_l+i\omega_1)(E_l-E_m+i\omega_2)(E_m-E_n-i\omega_3)},\label{lehmann}
\end{align}
where $|l\rangle$ and $E_l$ are the $l$th lowest eigenvalue and corresponding eigenvector of the Hamiltonian; $c^\dagger_i$ and $c_i$ represent creation and annihilation operators of an electron on the impurity site, respectively and index $i$ $(i=1,2,3)$ of these operators and also those of $g_{12}$ and $\chi_{1234}$ later appeared in this section are shorthand notation of combined index for the spin $\sigma_i$ and orbital $o_i$ degree of freedom, e.g., $c_1\equiv c_{\sigma_1,o_1}$; $\omega_i$  denote the fermionic Matsubara frequency: $\omega_i=\pi(2n_i+1)/\beta$.
Since these terms contain the factors in the denominator with the difference of two eigenvalues, e.g., $E_m-E_n$, terms with nearly degenerated two high-energy eigenvectors $E_m$ and $E_n$ can have large contribution\cite{HHafermann2010}. Therefore the precise eigenvalues of whole energy range are necessary. Since the Lanczos ED is accurate only for low-energy eigenvectors, an alternative expression for the two-body Green's function is desired.

On the other hands, for the Fourier transform of the one-body Green's function $g_{12}(\tau)\equiv -\langle  {\cal T}[c_1(\tau)c^{\dagger}_2(0)]\rangle$ the expression with resolvents
\begin{align}
g_{12}(i\omega)&=
\frac{1}{Z}\sum_{l,m}e^{-\beta E_l}\Big\{ \frac{\langle l |c_1|m\rangle\langle m| c^\dagger_2 |l \rangle }{i\omega + E_l-E_m} \nonumber \\
&~~~~~~~~~~~~~~~~~~~+\frac{\langle l |c^\dagger_2|m\rangle\langle m| c_1|l\rangle }{i\omega +E_m-E_l}\Big\} \label{g1a}\\
&=\frac{1}{Z}\sum_le^{-\beta E_l}\Big\{ \langle l |c_1  \frac{1}{i\omega + E_l-{\cal H}}c^\dagger_2 |l \rangle \nonumber \\
&~~~~~~~~~~~~~~~~~-\langle l |c^\dagger_2  \frac{1}{-i\omega + E_l-{\cal H}} c_1 |l \rangle\Big\}\label{g1b}
\end{align}
is applicable to the Lanczos ED technique and has already adopted in the DMFT studies\cite{AGeorges1996,MCapone}.
Since eigenvector $|l\rangle$ only within the energy range of the thermal excitation contribute due to the presence of the Boltzmann factor $e^{-\beta E_l}$, $|l\rangle$ and $E_l$ can be accurately calculated at low temperatures with the Lanczos ED technique. The resolvents in Eq.~(\ref{g1b}) can be transformed into continued fractions using the Lanczos algorithm and this continued fraction can be terminated typically several hundreds floors to obtain accurate results even for systems with $\sim 10^7$ basis functions. This technique is called the recursion method\cite{VHeine1980}. The procedure is also equivalent to replacing $E_m$ and $|m\rangle$ in Eq.~(\ref{g1a}) by those approximately obtained within a subspace spanned by the Lanczos vectors, i.e., the Krylov subspace\cite{ZBai2000}.

The expression of the Fourier transform of the two-body Green's function
\begin{align}
\chi_{1234}(\tau_1,\tau_2;\tau_3,\tau_4)\equiv 
\langle {\cal T}[c_1(\tau_1)c_2(\tau_2)c_3^\dagger(\tau_3)c_4^\dagger(\tau_4)]\rangle, \label{g2}
\end{align}
presented here consists of terms each of which has three or two resolvents in the form $1/(i\omega+E_l-{\cal H})$ and the factor $e^{-\beta E_l}$ sharing the same eigenvalue $E_l$. Because of this feature, only terms with $E_l$ within the energy range of the thermal excitation contribute similar to Eq.~(\ref{g1b}) of the one-body Green's function. Therefore, unlike the Lehmann representation, the denominators of these terms are always large with the high-energy eigenvectors of ${\cal H}$ and thus no significant contribution of high-energy eigenvectors is expected. This makes it possible to approximate the eigenvectors of ${\cal H}$ by those calculated within the Krylov subspace constructed by the Lanczos algorithm, which is less accurate in high-energy eigenvectors.

The detailed derivation of the expression is in Appendix A. The expression can be separated into three components in terms of the similarity to the three scattering channels: those of the horizontal $\chi^{\rm ph}$, and vertical $\chi^{\underline{{\rm ph}}}$ particle-hole, and the particle-particle $\chi^{\rm pp}$ types: 
\begin{align}
\chi_{1234}=\chi^{\rm ph}_{1234}+\chi^{\underline{{\rm ph}}}_{1234}+\chi^{\rm pp}_{1234}. \label{g2omega}
\end{align}
Each of the components can be given as
\begin{widetext}
\begin{align}
&\chi^{\rm ph}_{1234}(i\omega_1,i\omega_2;i\omega_3,i\omega_4)= \nonumber\\
&-\frac{1}{Z}\sum_l e^{-\beta E_l}\langle l |\big(
(c_1 \| c_4^\dagger)_{E_l+i\omega_1}-(c_4^\dagger \| c_1)_{E_l-i\omega_4}\big\|
(c_2 \| c_3^\dagger)_{E_l+i\omega_3}-(c_3^\dagger \| c_2)_{E_l-i\omega_2}
\big)_{ E_l-i(\omega_2-\omega_3)}| l \rangle \nonumber\\
&-\frac{1}{Z}\sum_l e^{-\beta E_l}\langle l |\big(
(c_2 \| c_3^\dagger)_{E_l+i\omega_2}-(c_3^\dagger \| c_2)_{E_l-i\omega_3}\big\|
(c_1 \| c_4^\dagger)_{E_l+i\omega_4}-(c_4^\dagger \| c_1)_{E_l-i\omega_1}
\big)_{E_l+i(\omega_2-\omega_3)  }| l \rangle \nonumber\\
&+\frac{1}{Z}\sum_{l,m}\delta_{E_l,E_m}e^{-\beta E_l}
\langle l |\big[ 
(c_1 \| c_4^\dagger)_{E_l+i\omega_1}-(c_4^\dagger \| c_1)_{E_l-i\omega_4}
\big] | m \rangle \langle m | \big[ 
(c_2 \bigtriangleup c_3^\dagger)_{\substack{E_l+i\omega_2 \\ E_l+i\omega_3}}-(c_3^\dagger \bigtriangleup c_2)_{\substack{E_l -i\omega_3\\ E_l-i\omega_2}}
\big]  | l \rangle \nonumber\\
&+\frac{1}{Z}\sum_{l,m}\delta_{E_l,E_m}e^{-\beta E_l}
  \langle l |\big[ 
(c_1 \bigtriangleup c_4^\dagger)_{\substack{E_l+ i\omega_1\\E_l+ i\omega_4}}-(c_4^\dagger \bigtriangleup c_1)_{\substack{E_l-i\omega_4 \\ E_l-i\omega_1}}
\big] | m \rangle\langle m |\big[ 
(c_2 \| c_3^\dagger)_{E_l+i\omega_2}-(c_3^\dagger \| c_2)_{E_l-i\omega_3}
\big]  | l \rangle \nonumber\\
&+\beta\delta_{\omega_1,\omega_4}\frac{1}{Z}\sum_{l,m}\delta_{E_l,E_m}e^{-\beta E_l}
\langle l |\big[ 
(c_1 \| c_4^\dagger)_{E_l+i\omega_1}-(c_4^\dagger \| c_1)_{E_l-i\omega_1} 
\big] | m \rangle \langle m | \big[ 
(c_2 \| c_3^\dagger)_{E_l+i\omega_2}-(c_3^\dagger \| c_2)_{E_l-i\omega_2}
\big]  | l \rangle  \label{g2ph1}
\end{align}
\begin{align}
&\chi^{\underline{{\rm ph}}}_{1234}(i\omega_1,i\omega_2;i\omega_3,i\omega_4)= \nonumber\\
&+\frac{1}{Z}\sum_l e^{-\beta E_l}\langle l |\big( 
(c_1 \| c_3^\dagger)_{E_l+i\omega_1}-(c_3^\dagger \| c_1)_{E_l-i\omega_3}\big\|
(c_2 \| c_4^\dagger)_{E_l+i\omega_4}-(c_4^\dagger \| c_2)_{E_l-i\omega_2}
\big)_{E_l+i(\omega_1-\omega_3)}| l \rangle \nonumber\\
&+\frac{1}{Z}\sum_l e^{-\beta E_l}\langle l |\big(
(c_2 \| c_4^\dagger)_{E_l+i\omega_2}-(c_4^\dagger \| c_2)_{E_l-i\omega_4}\big\|
(c_1 \| c_3^\dagger)_{E_l+i\omega_3}-(c_3^\dagger \| c_1)_{E_l-i\omega_1}
\big)_{E_l -i(\omega_1-\omega_3) }| l \rangle \nonumber\\
&-\frac{1}{Z}\sum_{l,m}\delta_{E_l,E_m}e^{-\beta E_l} 
\langle l | \big[ 
(c_1 \| c_3^\dagger)_{E_l+i\omega_1}-(c_3^\dagger \| c_1)_{E_l-i\omega_3}
\big] | m \rangle \langle m | \big[ 
(c_2 \bigtriangleup c_4^\dagger)_{\substack{E_l+ i\omega_2\\ E_l+i\omega_4}}-(c_4^\dagger \bigtriangleup c_2)_{\substack{ E_l-i\omega_4 \\ E_l-i\omega_2}}
\big]  | l \rangle \nonumber\\
&-\frac{1}{Z}\sum_{l,m}\delta_{E_l,E_m}e^{-\beta E_l}
 \langle l | \big[ 
(c_1 \bigtriangleup c_3^\dagger)_{\substack{E_l+i\omega_1 \\ E_l+i\omega_3}}-(c_3^\dagger \bigtriangleup c_1)_{\substack{E_l- i\omega_3\\ E_l-i\omega_1}}
\big] | m \rangle\langle m |\big[ 
(c_2 \| c_4^\dagger)_{E_l+i\omega_2}-(c_4^\dagger \| c_2)_{E_l-i\omega_4}
\big]  | l \rangle \nonumber\\
&-\beta\delta_{\omega_1,\omega_3}\frac{1}{Z}\sum_{l,m}\delta_{E_l,E_m}e^{-\beta E_l}
\langle l |\big[ 
(c_1 \| c_3^\dagger)_{E_l+i\omega_1}-(c_3^\dagger \| c_1)_{E_l-i\omega_1} 
\big] | m \rangle \langle m | \big[ 
(c_2 \| c_4^\dagger)_{E_l+i\omega_2}-(c_4^\dagger \| c_2)_{E_l-i\omega_2}
\big]  | l \rangle \label{g2ph2}
\end{align}
\begin{align}
&\chi^{\rm pp}_{1234}(i\omega_1,i\omega_2;i\omega_3,i\omega_4)= \nonumber\\
&-\frac{1}{Z}\sum_l e^{-\beta E_l}\langle l |\big(
(c_1 \| c_2)_{E_l+i\omega_1}-(c_2 \| c_1)_{E_l+i\omega_2}\big\|
 (c_3^\dagger \| c_4^\dagger)_{E_l+i\omega_4}-(c_4^\dagger \| c_3^\dagger)_{E_l+i\omega_3}
 \big)_{ E_l+i(\omega_1+\omega_2)}| l \rangle \nonumber\\
&-\frac{1}{Z}\sum_l e^{-\beta E_l}\langle l |\big( 
(c_3^\dagger \| c_4^\dagger)_{E_l-i\omega_3}-(c_4^\dagger \| c_3^\dagger)_{E_l-i\omega_4}\big\|
(c_1 \| c_2)_{E_l-i\omega_2}-(c_2 \| c_1)_{E_l-i\omega_1}
\big)_{E_l- i(\omega_1+\omega_2)}| l \rangle \nonumber \\
&+\frac{1}{Z}\sum_{l,m}\delta_{E_l,E_m}e^{-\beta E_l}
\langle l |\big[ 
(c_1 \| c_2)_{E_l+i\omega_1}-(c_2\| c_1)_{E_l+i\omega_2}
\big] | m \rangle \langle m | \big[ 
(c_3^\dagger \bigtriangleup c_4^\dagger)_{\substack{E_l-i\omega_3 \\E_l+i\omega_4}}-(c_4^\dagger \bigtriangleup c_3^\dagger )_{\substack{E_l-i\omega_4\\E_l +i\omega_3}}
\big] \big| l \big\rangle \nonumber\\
&+\frac{1}{Z}\sum_{l,m}\delta_{E_l,E_m}e^{-\beta E_l}
 \langle l | \big[ 
(c_1 \bigtriangleup c_2)_{\substack{E_l+i\omega_1 \\ E_l-i\omega_2}}-(c_2 \bigtriangleup c_1)_{\substack{E_l+ i\omega_2\\ E_l -i\omega_1}}
\big] | m \rangle \langle m | \big[ 
(c_3^\dagger \| c_4^\dagger)_{E_l-i\omega_3}-(c_4^\dagger \| c_3^\dagger)_{E_l-i\omega_4}
\big]   | l\rangle\nonumber\\
&+\beta\delta_{\omega_1,-\omega_2}\frac{1}{Z}\sum_{l,m}\delta_{E_l,E_m}e^{-\beta E_l}
\langle l |\big[ 
(c_1 \| c_2)_{E_l+i\omega_1}-(c_2 \| c_1)_{E_l-i\omega_1} 
\big]  | m \rangle \langle m | \big[ 
(c_3^\dagger \| c_4^\dagger)_{E_l-i\omega_3}-(c_4^\dagger \| c_3^\dagger)_{E_l+i\omega_3}
\big]  | l \rangle \label{g2pp}
\end{align}
\end{widetext}
where, the operators in the form of $({\cal A}||{\cal B})_z$ and $({\cal A}\bigtriangleup{\cal B})_{\substack{z \\ z'}}$ are the abbreviations of those contain one and two resolvents, respectively, and are defined as
\begin{align}
({\cal A}||{\cal B})_z &\equiv{\cal A}\frac{1}{z  -{\cal H}'}{\cal B}, \label{rezo1}\\
({\cal A}\bigtriangleup{\cal B})_{\substack{z \\ z'}} &\equiv {\cal A}\frac{1}{z  -{\cal H}}\frac{1}{z'-{\cal H}}{\cal B}. \label{rezo2}
\end{align}
The resolvent with the Hamiltonian ${\cal H}'$ in the denominator of Eq.~(\ref{rezo1})  is the same to that of ${\cal H}$ except that all the eigenvectors $|l\rangle$ whose eigenvalue $E_l$ are equal to the real part of $z$ are projected out as
\begin{align}
({\cal A}||{\cal B})_z =\sum_{l,E_l\ne \textrm{Re}\,z}{\cal A}|l\rangle\frac{1}{z  -E_l}\langle l|{\cal B}.\label{resod}
\end{align}
The first two lines of the right-hand side of each of Eqs.~(\ref{g2ph1})-(\ref{g2pp}) contain 8 terms with three resolvents such as
\begin{align}
&-\sum_l \frac{e^{-\beta E_l}}{Z}\langle l |\big( 
(c_1 \| c_4^\dagger)_{E_l+i\omega_1}\big\|
(c_2 \| c_3^\dagger)_{E_l+i\omega_3}\big)_{E_l- i(\omega_2-\omega_3)}| l \rangle \nonumber \\
&=-\sum_l \frac{e^{-\beta E_l}}{Z}\langle l |c_1 \frac{1}{i\omega_1 + E_l-{\cal H}} c_4^\dagger \nonumber \\
&~~~~~\times \frac{1}{i(\omega_3-\omega_2) + E_l-{\cal H}'}c_2 \frac{1}{i\omega_3 + E_l-{\cal H}}  c_3^\dagger| l \rangle.\label{major}
\end{align}
In total 4!\,(=24) of terms of this kind exist and we call them the major terms. Each term contains a resolvent with a bosonic Matsubara frequency, e.g., that with $i(\omega_3-\omega_2)$ in Eq.~(\ref{major}), and for this resolvent, eigenvectors with eigenvalue $E_l$ are projected out to avoid divergence at zero frequency. The proper treatment of these special cases further requires 36 counter terms (for details, see Appendix A) and there are two kinds of them: one is those consist of products of two factors containing one and two resolvents as
\begin{align}
&\sum_{\substack{ l,m \\  E_l=E_m}}\!\!\frac{e^{-\beta E_l}}{Z} 
\langle l | (c_1 \| c_4^\dagger)_{E_l+i\omega_1} | m \rangle \langle m | 
(c_2 \bigtriangleup c_3^\dagger)_{\substack{E_l+ i\omega_2\\ E_l+i\omega_3}}  | l \rangle \nonumber\\
&=\sum_{\substack{ l,m \\  E_l=E_m}}\!\!\frac{e^{-\beta E_l}}{Z} \langle l |c_1 \frac{1}{i\omega_1 + E_l-{\cal H}} c_4^\dagger | m \rangle   \nonumber\\
&~~~~~~~~~~~~\times \langle m |c_2\frac{1}{i\omega_2 + E_l-{\cal H}}\frac{1}{i\omega_3 + E_l-{\cal H}}  c_3^\dagger| l \rangle.\label{minor1}
\end{align}
and the other kind of the counter terms have the form
\begin{align}
&\beta\delta_{\omega_1,\omega_4}\!\!\!\!\!\!\sum_{\substack{ l,m \\  E_l=E_m}}\!\!\!\!\!\frac{e^{-\beta E_l}}{Z}
\langle l| (c_1 \| c_4^\dagger)_{E_l+i\omega_1} | m \rangle \langle m | 
(c_2 \| c_3^\dagger)_{E_l+i\omega_2}  | l \rangle \nonumber\\
&=\beta\delta_{\omega_1,\omega_4}\!\!\!\sum_{\substack{ l,m \\  E_l=E_m}}\!\!\!\frac{e^{-\beta E_l}}{Z}
\langle l |c_1 \frac{1}{i\omega_1 + E_l-{\cal H}} c_4^\dagger | m \rangle   \nonumber\\
&~~~~~~~~~~~~~~~~~~~~~~~~~~~~\times \langle m | c_2  \frac{1}{i\omega_2 + E_l-{\cal H}}  c_3^\dagger |l\rangle.\label{minor2}
\end{align}
Note that if the eigenenergies of wave functions with different electron numbers are degenerated, the counter terms in Eq.~(\ref{g2pp}) can have non-zero values even \textit{without} finite superconducting order parameter. This can happen if the system possesses the electron-hole symmetry, e.g., half-filled square-lattice Hubbard model with the nearest neighbor hopping in this study.

\section{Approximation of two-body Green's function with Lanczos algorithm\label{LanczosED}}
Having the new expression in hand,  in this section we discuss how to calculate the two-body Green's function approximately with the Lanczos algorithm\cite{ZBai2000}.
The Lanczos algorithm is a unitary transformation, which converts a symmetric or Hermitian matrix ${\cal H}$ into a tridiagonal form:
\begin{align}
T^{(n)}=\left(\begin{array}{ccccc}
a_1 & b_1 &  0 & \cdots &  0\\
b_1 & a_2 & b_2  & \ddots  &\vdots  \\
0     & b_2 & a_3 & \ddots & 0\\
\vdots &\ddots   &  \ddots&  \ddots& b_{n-1}\\
0  &  \cdots   & 0 & b_{n-1} & a_n
\end{array}\right).
\end{align}
Starting from a properly chosen initial vector $|v_1\rangle$, it creates one of the orthonormal basis vector $|v_n\rangle$ in every iteration step,  and at $n$th step these basis vectors span the Krylov subspace ${\cal K}^n(|v_1\rangle,{\cal H})$=span$\{|v_1\rangle,$ ${\cal H}|v_1\rangle,$ ${\cal H}^2|v_1\rangle,\dots,$ ${\cal H}^{n-1}|v_1\rangle\}$. In practice, because of round off error, the orthogonality of the vectors $|v_n\rangle$ breaks midway through the iteration. This occurs as soon as the lowest eigenvector converges and one may set the criterion to terminate the iteration when this convergence is reached:
\begin{align}
|b_{n-1}s_{1,n}| < \varepsilon_{\rm lan},
\end{align}
where $s_{i,j}$ is the $j$th component of the eigenvector of $T^{(n)}$ with the $i$th lowest eigenvalue $\theta_i$, i.e., $T^{(n)}|s_i\rangle=\theta_i|s_i\rangle$.  In this study, the threshold $\varepsilon_{\rm lan}=\alpha_{\rm lan}\sqrt{N_{\rm sys}}\varepsilon$ is assumed, where $N_{\rm sys}$ is the order of matrix ${\cal H}$, $\varepsilon$ denotes the machine accuracy $\varepsilon=10^{-15}$ and $\alpha_{\rm lan}=$10. The $i$th lowest eigenvalue $E_i$ and corresponding eigenvector $|i\rangle$ of ${\cal H}$ can be approximated as
\begin{align}
|i\rangle\approx \sum_{j=1}^n s_{i,j}|v_j\rangle,~~~E_i\approx \theta_i. \nonumber
\end{align}
High accuracy (typically more than 10 digits) can be expected for the eigenvector $|1\rangle$ of the lowest eigenvalue $E_1$ and the rest of them are less accurate. 

Now, let us consider the major term with the form
\begin{align}
\sum_l \frac{e^{-\beta E_l}}{Z} &\langle l |{\cal O}_1 \frac{1}{i\omega + E_l-{\cal H}} {\cal O}_2\nonumber \\
 &\times  \frac{1}{i\nu + E_l-{\cal H}'}{\cal O}_3  \frac{1}{i\omega' + E_l-{\cal H}}  {\cal O}_4| l \rangle,  \label{major1}
\end{align}
 where each ${\cal O}_i$ $(i=1,2,3,4)$ is one of $c_1$, $c_2$, $c^\dagger_3$ or $c^\dagger_4$ and  $\nu$ is the bosonic and $\omega$, and $\omega'$ are the fermionic Matsubara frequencies. The eigenvectors $|l\rangle$ need to be included in the calculations are limited to those of low energies by the Boltzmann factor $e^{-\beta E_l}$. Note that high accuracy is particularly needed for $E_l$ and $|l\rangle$ since it affects later calculations. Therefore, instead of calculating all the $E_l$ and $|l\rangle$ required at a time by the Lanczos algorithm describe above, it is preferable to use the so called restart Lanczos method\cite{ZBai2000}, where only one lowest eigenvector is calculated at a time and repeat the same Lanczos procedure except in each step $|v_n\rangle$ is orthogonalized to all the previously obtained eigenvectors. In this way, all the required $E_l$ and $|l\rangle$ can be calculated with high precision.

 One of the three resolvents in Eq.~(\ref{major1}) on the left can be obtained approximately using $N_{\rm L}$ eigenvalues $E_m^{\rm L}$ and eigenvectors $|m^{\rm L}\rangle$ generated by combined use of the restart Lanczos method for low lying eigenvectors within the reach of the thermal excitation and the ordinary Lanczos method with the initial vector ${\cal O}_1^\dagger| l \rangle$ for rest of high-energy eigenvectors as
\begin{align}
\frac{1}{i\omega+E_l-{\cal H}}\approx \sum_{m=1}^{N_{\rm L}} |m^{\rm L}\rangle \frac{1}{i\omega+E_l-E_m^{\rm L}}\langle m^{\rm L}|.
\end{align}
The same can be done for the resolvent on the right using $N_{\rm R}$ eigenvalues $E_m^{\rm R}$ and eigenvectors $|m^{\rm R}\rangle$ generated by these Lanczos methods with the initial vector ${\cal O}_4| l \rangle$. Although the high-energy eigenvectors obtained with the ordinary Lanczos method are less accurate compared to the low-energy eigenvectors calculated with the restart Lanczos method, the resultant left and right resolvents have proper asymptotic behavior. For instance, 
\begin{align}
 \frac{1}{z-{\cal H}}{\cal O}_4| l \rangle =\sum_{n=0}^\infty \frac{1}{z^{n+1}}{\cal H}^n{\cal O}_4| l \rangle 
\end{align}
for the right resolvent. Obviously the expansion is correct up to the $N_{\rm R}$th order, since their coefficients belong to the Krylov space ${\cal K}^{N_{\rm R}}({\cal O}_4| l \rangle,{\cal H})$.

For the resolvent in the center, excitations through the left and right resolvents with different energy scales are required to be considered. To do so, vectors which represent excitation at the left $|v^L_\alpha\rangle$ and right $|v^R_\alpha\rangle$ resolvents with the energy $\Omega_\alpha$ and an artificial lifetime width $\gamma_\alpha$ are introduced as
\begin{align}
& |v^L_\alpha\rangle =\sum_{m=1}^{N_{\rm L}} {\cal O}_2^\dagger|m^{\rm L}\rangle \frac{\langle m^{\rm L} |{\cal O}_1^\dagger|l\rangle}{(\Omega_\alpha+E_l-E_m^{\rm L})^2+\gamma_\alpha^2},\nonumber \\
 & |v^R_\alpha\rangle =\sum_{m=1}^{N_{\rm R}} {\cal O}_3|m^{\rm R}\rangle \frac{\langle m^{\rm R} |{\cal O}_4|l\rangle}{(\Omega_\alpha+E_l-E_m^{\rm R})^2+\gamma_\alpha^2}.  \label{exvectors}
\end{align}
We use these vectors with a finite number of reference energy points $\Omega_\alpha$ ($\alpha=1,2,\cdots,N_\alpha$) as the initial vectors of the band Lanczos method\cite{ZBai2000}, which is an extension of the Lanczos method with multiple initial vectors, to generate the basis set to construct approximated resolvent.
The Lanczos vectors need to be orthogonalized to the eigenvector $|l\rangle$ and, if exist, all the degenerated eigenvectors with the eigenvalue $E_l$ [see Eq.~(\ref{resod})]. It is, however, recommendable to calculate the low-energy eigenvectors within the reach of the thermal excitation by the restart Lanczos method and use the band Lanczos method to generate high-energy eigenvectors by orthogonalizing its initial and Lanczos vectors to the former. It is essential to choose one of $\Omega_\alpha$ to be $\Omega_\alpha \gg E_{\rm max}-E_l$ to make $\chi_{1234}$ have proper asymptotic behavior, where $E_{\rm max}$ is the maximum eigenvalue of ${\cal H}$.

As demonstrated in Appendix C, the resultant $\chi_{1234}$ (or the vertex function) converges rapidly as the number of the reference energy points $\Omega_\alpha$ increases. In the calculations of the 2D Hubbard model in this study we adopted four reference energy points $\Omega_1=0$, $\Omega_2=0.02W$,  $\Omega_3=0.04W$ and $\Omega_4=4W$ with all $\gamma_\alpha=0.1W$, where the effective band width $W$ is defined as 
 \begin{align}
 W \equiv \sqrt{U^2+64t^2}\label{W}. 
 \end{align}
To further reduce the burden of the computational tasks, the initial vectors $|v^L_\alpha\rangle$ and $|v^R_\alpha\rangle$ of the four major terms in each of the first two lines on the right side of Eqs.~(\ref{g2ph1})-(\ref{g2pp}) can be combined (they are not necessarily linearly independent and the number of required initial vectors is smaller than it appears) and the basis set generated by the band Lanczos method with the combined initial vectors can be shared in the calculation of these four major terms.

Once $N_{\rm C}$ eigenvalues $E_n^{\rm C}$ and eigenvectors $|n^{\rm C}\rangle$ are obtained with the band Lanczos method, we can approximate the major term in the form of Eq.~(\ref{major1}) as
\begin{widetext}
\begin{align}
\frac{1}{Z}\sum_{l=1}^{N_{\rm I}} \sum_{m=1}^{N_{\rm L}}\sum_{n=1}^{N_{\rm C}}\sum_{m'=1}^{N_{\rm R}}  e^{-\beta E_l}\frac{\langle l |{\cal O}_1|m^{\rm L}\rangle\langle m^{\rm L} |{\cal O}_2|n^{\rm C} \rangle\langle n^{\rm C}|{\cal O}_3|{m'}^{\rm R}\rangle\langle {m'}^{\rm R} |{\cal O}_4| l \rangle }{(i\omega + E_l-E_m^{\rm L})(i\nu + E_l-E_n^{\rm C})(i\omega' + E_l-E_{m'}^{\rm R})},  \label{major2}
\end{align}
\end{widetext}
which bears a resemblance to Eq.~(\ref{lehmann}) but all the three factors in the denominator and $e^{-\beta E_l}$ shares the same $E_l$. This makes terms with all the eigenvalues $E_l$, $E^{\rm L}_m$, $E^{\rm C}_n$ and $E^{\rm R}_{m'}$ being within the energy range of thermal excitation mostly contribute and for these terms, one can use the restart Lanczos method to obtain accurate eigenvectors and eigenvalues. For the rest of terms with high-energy excitations, the combined use of the ordinary and band Lanczos methods as described ensures accurate asymptotic behavior. Since the typical number of the Lanczos vectors, $N_{\rm L}$, $N_{\rm C}$, and $N_{\rm R}$, is several hundreds even for $10^6$ actual basis functions, the method proposed here renders drastic reduction of computational workload over ordinary ED method.

The counter terms, such as Eqs.~(\ref{minor1}) and (\ref{minor2}) can be calculated in the similar way. For instance, the second factor with two resolvents in Eq.~(\ref{minor1}) can be approximated using $N_{\rm R}$ eigenvalues $E_n^{\rm R}$ and eigenvectors $|n^{\rm R}\rangle$ generated from the band Lanczos algorithm with the initial vectors $c^\dagger_3|l\rangle$ and $c^\dagger_2|m\rangle$ for all the degenerated  eigenvectors $|l\rangle$ and $|m\rangle$ ($E_l=E_m$) as
\begin{align}
\sum_{n=1}^{N_{\rm R}}  \frac{\langle m |c_2|n^{\rm R}\rangle\langle n^{\rm R} |c^\dagger_3| l \rangle }{(i\omega_2 + E_l-E_n^{\rm R})(i\omega_3 + E_l-E_n^{\rm R})}.
\end{align}

\section{Ladder Dual Fermion Approximation\label{LDFA}}
In this section, a brief overview of LDFA\cite{ANRubtsov2008,ANRubtsov2009,HHafermann2009} and some technical points specific to the present calculation scheme are provided. 
The action for the 2D Hubbard model on the square lattice with the nearest-neighbor hopping integral $t$ and the on-site Coulomb interaction $U$ is
\begin{align}
S[\overline{c},c]=&-\sum_{\bm{k}\omega\sigma}\overline{c}_{\bm{k}\omega\sigma}(i\omega + \mu -\varepsilon_{\bm{k}})c_{\bm{k}\omega\sigma} \nonumber\\
&~~~~~~+U\sum_i\int_0^{\beta}d\tau\,\overline{c}_{i\tau\uparrow}c_{i\tau\uparrow}\overline{c}_{i\tau\downarrow}c_{i\tau\downarrow},\label{S2DHub}
\end{align}
where $\overline{c}_{\bm{k}\omega\sigma}$ ($c_{\bm{k}\omega\sigma}$) and  $\overline{c}_{i\tau\sigma}$ ($c_{i\tau\sigma}$) are the fermionic Grassmann fields corresponding to the creation (annihilation) operators $c^{\dagger}_{\bm{k}\sigma}$ ($c_{\bm{k}\sigma}$) and $c^{\dagger}_{i\sigma}$ ($c_{i\sigma}$), respectively; $\omega$ represents the fermionic Matsubara frequency, $\mu$ denotes the chemical potential, and $\varepsilon_{\bm{k}}=-2t(\cos k_x+\cos k_y)$. 

The IAM at site $i$ can be written as 
\begin{align}
S_{\rm imp}[\overline{c}_i,c_i]&=-\sum_{\omega\sigma}\overline{c}_{i\omega\sigma}(i\omega + \mu -\Delta_{\omega})c_{i\omega\sigma}\nonumber \\
&~~~~~~+U\int_0^{\beta}d\tau\,\overline{c}_{i\tau\uparrow}c_{i\tau\uparrow}\overline{c}_{i\tau\downarrow}c_{i\tau\downarrow},\label{SIAM}
\end{align}
where $\Delta_\omega$ denotes the hybridization function, which is arbitrary at this point.
The lattice action in Eq.~(\ref{S2DHub}) can be represented by the action of the IAM for each site $i$ plus a correction term:
\begin{align}
S[\overline{c},c]=\sum_i S_{\rm imp}[\overline{c}_i,c_i] + \sum_{\bm{k}\omega\sigma}\overline{c}_{\bm{k}\omega\sigma}(\varepsilon_{\bm{k}}-\Delta_\omega)c_{\bm{k}\omega\sigma}.\label{Scc}
\end{align}

Instead of directly performing the perturbative calculations with Eq.\,(\ref{Scc}), a new fermionic auxiliary field, which is called the dual fermion, $f_{\bm{k}\omega\sigma}$ is introduced using a Hubbard-Stratonovich transformation\cite{ANRubtsov2008,ANRubtsov2009}.  The original action can be mapped onto that of the dual fermion by integrating out the real electron field $c_{\bm{k}\omega\sigma}$. In this way, one can separate the problem of solving the IAM to obtain the local approximation and the perturbative corrections for the spatial correlations, avoiding the double counting of local contributions. The action of $f_{\bm{k}\omega\sigma}$ within the fourth order is 
\begin{align}
S_d[\overline{f},f]&=-\sum_{\bm{k}\omega\sigma}\overline{f}_{\bm{k}\omega\sigma}[G^{d,0}_{\bm{k}\omega}]^{-1}f_{\bm{k}\omega\sigma} \nonumber \\
&~~~~~~~~~~~~~~~~-\frac{1}{4}\sum_{\substack{1234 \\ i}}\gamma^{(4)}_{1234}\overline{f}_{i1}\overline{f}_{i2}f_{i3}f_{i4},
\end{align}
where shorthand notations such as $1\equiv(\omega_1,\sigma_1)$ are used for the indices; $G^{d,0}_{\bm{k}\omega}$ denotes non-interacting dual-fermion one-body Green's function, and $\gamma^{(4)}_{1234}$ represents  the reducible four-point vertex function of the impurity site for original electrons and they are defined using the local one-body $g_\omega$ and two-body $\chi_{1234}$ Green's functions  of the impurity site for original electrons as
\begin{align}
G^{d,0}_{\bm{k}\omega}&=-g_{\omega}+\left[g_{\omega}^{-1}+\Delta_{\omega}-\varepsilon_{\bm{k}}\right]^{-1}, \label{Gd0}\\
\gamma^{(4)}_{1234}&=g^{-1}_1g^{-1}_2\left[\chi_{1234}-\beta(\delta_{14}\delta_{23}-\delta_{13}\delta_{24})g_1g_2\right]g^{-1}_3g^{-1}_4.\label{vtx4def}
\end{align}
Note that diagrams containing the six-point vertex function give a negligible contribution\cite{HHafermann2009}.
For the sake of convenience, we use notation 
\begin{align}
\gamma^{\sigma_1\sigma_2\sigma_3\sigma_4}_{\omega\omega';\Omega}\equiv\gamma^{(4)}_{(\omega,\sigma_1),(\omega'+\Omega,\sigma_2),(\omega',\sigma_3),(\omega+\Omega,\sigma_4)}.\label{vtx4note}
\end{align}
Since we are dealing with the PM state, the system has spin rotational symmetry and the vertex function can be diagonalized with respect to the spin indices and separated into the charge ($S=0$) and spin ($S=1$) components:
\begin{align}
\gamma^{({\rm ch})}_{\omega\omega';\Omega}&= \gamma^{\uparrow\uparrow\uparrow\uparrow}_{\omega\omega';\Omega}+\gamma^{\uparrow\downarrow\downarrow\uparrow}_{\omega\omega';\Omega}, \\
\gamma^{({\rm sp})}_{\omega\omega';\Omega}&= \gamma^{\uparrow\uparrow\uparrow\uparrow}_{\omega\omega';\Omega}-\gamma^{\uparrow\downarrow\downarrow\uparrow}_{\omega\omega';\Omega}=\gamma^{\uparrow\downarrow\uparrow\downarrow}_{\omega\omega';\Omega}=\gamma^{\downarrow\uparrow\downarrow\uparrow}_{\omega\omega';\Omega}.
\end{align}

To include effects of long-range spin fluctuations, the ladder diagram of the particle-hole channel is taken into account, which is considered to be the dominant correction to DMFT for the spatial fluctuations in low temperatures at half filling.
The Bethe-Salpeter equation of the dual fermion for the charge ($\alpha=$ch) and spin ($\alpha=$sp) components are 
\begin{align}
\Gamma^{(\alpha)}_{\omega\omega';\bm{q}\Omega}=\gamma^{(\alpha)}_{\omega\omega';\Omega}+\frac{1}{\beta}\sum_{\omega''}\gamma^{(\alpha)}_{\omega\omega'';\Omega}\chi^{d,0}_{\bm{q}\omega''\Omega}\Gamma^{(\alpha)}_{\omega''\omega';\bm{q}\Omega},\label{BSE}
\end{align}
where
\begin{align}
\chi^{d,0}_{\bm{q}\omega\Omega}=-\frac{1}{N}\sum_{\bm{k}}G^d_{\bm{k}\omega}G^d_{\bm{k}+\bm{q},\omega+\Omega}.
\end{align}
We define the effective interaction of each component $\alpha$ as
\begin{align}
{\cal V}^{(\alpha)}_{\omega;\bm{q}\Omega}=\frac{1}{2\beta}\sum_{\omega'}\gamma^{(\alpha)}_{\omega\omega';\Omega}\chi^{d,0}_{\bm{q}\omega'\Omega} \left(\Gamma^{(\alpha)}_{\omega'\omega;\bm{q}\Omega}-\frac{1}{2}\gamma^{(\alpha)}_{\omega'\omega;\Omega}\right) \label{sigma2}
\end{align}
and the self-energy for the dual fermion can be written as 
\begin{align}
\Sigma^d_{\bm{k}\omega}&=\frac{-1}{\beta N}\sum_{\bm{q}\omega'}\gamma^{({\rm ch})}_{\omega\omega';0}G^d_{\bm{q}\omega'} \nonumber\\
&+\frac{1}{\beta N}\sum_{\bm{q}\Omega}\left({\cal V}^{({\rm ch})}_{\omega;\bm{q}\Omega}+3{\cal V}^{({\rm sp})}_{\omega;\bm{q}\Omega}\right)G^d_{\bm{k}+\bm{q},\omega+\Omega}.\label{sigmad}
\end{align}
The Green's function of the dual fermion is obtained from Dyson's equation
\begin{align}
[G^{d}_{\bm{k}\omega}]^{-1}=[G^{d,0}_{\bm{k}\omega}]^{-1}-\Sigma^d_{\bm{k}\omega}.\label{DEd}
\end{align}
The electron Green's function can be obtained from its dual counterpart as
\begin{align}
G_{\bm{k}\omega}=&[\varepsilon_{\bm{k}}-\Delta_{\omega}]^{-1}g_{\omega}^{-1} G^d_{\bm{k}\omega}g_{\omega}^{-1}[\varepsilon_{\bm{k}} -\Delta_{\omega}]^{-1}\nonumber \\
                                &-[\varepsilon_{\bm{k}}-\Delta_{\omega}]^{-1}.
\end{align}
There is also similar one-to-one correspondence between electrons and their dual counterparts for the higher-order correlation functions\cite{SBrener2008,ANRubtsov2009}.

The local corrections can be efficiently included in the impurity problem with a proper choice of $\Delta_{\omega}$. For the purpose, the condition $\langle G^d_{\bm{k}\omega}\rangle_{\bm{k}}=0$, with which diagrams containing a local loop vanish, is commonly used, where $\langle \cdots\rangle_{\bm{k}}=(1/N)\sum_{\bm{k}}\cdots$. This condition is reduced to the self-consistent condition of DMFT for the non-interacting dual fermions and therefore DMFT can be regarded as the lowest order in DFA\cite{ANRubtsov2009}. The nonlocal corrections can be included in $\Sigma^d_{\bm{k}\omega}$ by higher orders of dual fermion perturbation theory as already discussed.

In the calculations of the dual fermionic quantities in Eqs.\,(\ref{BSE})-(\ref{DEd}), the frequency cutoff $-N_{\omega}+1\leq n \leq N_{\omega}$ is set for the variables with the fermionic Matsubara frequency $\omega_n=(2n-1)\pi/\beta$ and $|m|\leq 2N_{\omega}-1$ for those with the bosonic Matsubara frequency $\Omega_m=2m\pi/\beta$. Choosing $N_{\omega}\approx W\beta/\pi$ is found to be sufficient to obtain accurate results, where $W$ is the effective band width in Eq.\,(\ref{W}). 

\begin{figure}
\includegraphics[width=8cm]{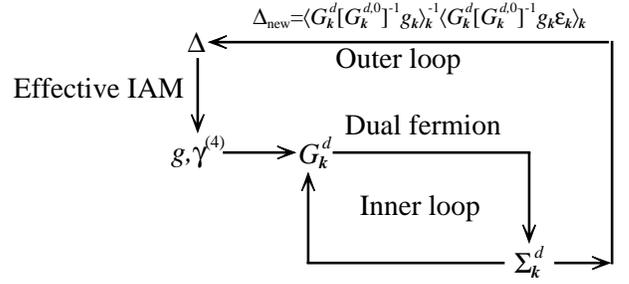}
\caption{\label{DFADiag}Illustration of the computational procedure of LDFA.}
\end{figure}
In the ED method, the conduction band of the effective IAM is replaced by discretized $N_{\rm b}$ energy levels (or bath sites) $l$ with energy $\varepsilon^{\rm b}_l$ and hybridization strength $V_l$ to the impurity orbital. The Hamiltonian can be written as 
\begin{align}
{\cal H}_{\rm imp}= \sum_{l\sigma}\left\{\varepsilon^{\rm b}_l a^\dagger_{l\sigma} a_{l\sigma}+V_l(a^\dagger_{l\sigma} c_{\sigma} +{\rm h.c.})\right\}+Un_{\uparrow}n_{\downarrow}, \label{IAM}
\end{align}
where $c^{\dagger}_{\sigma}$ ($c_{\sigma}$) and $a^{\dagger}_{l\sigma}$ ($a_{l\sigma}$) are the creation (annihilation) operators of an electron on the impurity site and bath site $l$, respectively, and $n_{\sigma}\equiv c^{\dagger}_{\sigma}c_{\sigma}$.
All the effects of the lattice and interaction except for the impurity site are encoded in the parameter set $\{\varepsilon^{\rm b}_l,V_l\}$ or the hybridization function
\begin{align}
\Delta(z;\{\varepsilon^{\rm b}_l,V_l\})= \sum_{l=1}^{N_{\rm b}}\frac{|V_l|^2}{z-\varepsilon^{\rm b}_l}. \label{Delta}
\end{align}

The outline of the computational procedure of LDFA is as follows.
The one-body Green's function $g_\omega$ at the impurity site can be obtained by the Lanczos ED method utilizing Eq.~(\ref{g1b}). 
The two-body Green's function $\chi_{1234}$ at the impurity site can be calculated using Eqs.~(\ref{g2})-(\ref{g2pp}) and the prescription in Sec.~\ref{LanczosED}. Once $g_\omega$ and $\chi_{1234}$ are obtained, the self-energy $\Sigma^d_{\bm{k}\omega}$ and Green's function $G^d_{\bm{k}\omega}$ of the dual fermion are calculated within the ladder approximation in the particle-hole channel using Eqs.\,(\ref{Gd0})-(\ref{DEd}).
$\Sigma^d_{\bm{k}\omega}$ and $G^d_{\bm{k}\omega}$ must be calculated iteratively until self-consistency is reached in the same manner to the fluctuation exchange approximation\cite{NEBickers1991} (the inner loop in Fig.~\ref{DFADiag}). More technical details can be found in Refs.~\onlinecite{HHafermann2010,JOtsuki2014}. The modified Broyden's method is applied to accelerate the convergence of the self-consistency loop of $\Sigma^d_{\bm{k}\omega}$\cite{RZitko}.

We further require the parameter set $\{\varepsilon^{\rm b}_l,V_l\}$ of the effective IAM to fulfill the condition $\langle G^d_{\bm{k}\omega}\rangle_{\bm{k}}=0$. To this end, we first choose the initial guess of the parameter set $\{\varepsilon^{\rm b}_l,V_l\}$, e.g., that of DMFT, and calculate $G^d_{\bm{k}\omega}$. We update the parameter set, calculate $G^d_{\bm{k}\omega}$ again and repeat this procedure until $\langle G^d_{\bm{k}\omega}\rangle_{\bm{k}}=0$ is fulfilled (the outer loop in Fig.~\ref{DFADiag}). 

For the update of the hybridization function, we use
\begin{align}
\Delta^{\rm new}_\omega=\left\langle G^d_{\bm{k}\omega}[G^{d,0}_{\bm{k}\omega}]^{-1}g_{\bm{k}\omega}\right\rangle^{-1}_{\bm{k}}\left\langle G^d_{\bm{k}\omega}[G^{d,0}_{\bm{k}\omega}]^{-1}g_{\bm{k}\omega}\varepsilon_{\bm{k}}\right\rangle_{\bm{k}}.\label{deltanewdfa}
\end{align}
The detailed derivation of Eq.\,(\ref{deltanewdfa}) can be found in Appendix B.
Once the new hybridization function is obtained, the parameter set $\{\varepsilon^{\rm b}_l,V_l\}$ for the next iteration is determined by minimizing the distant function defined as
\begin{align}
d=\sum_{p=1}^{2N_\omega}\frac{1}{|z_p|}\left|\Delta^{\rm new}(z_p)-\Delta(z_p;\{\varepsilon^{\rm b}_l,V_l\})\right|^2, 
\end{align}
where the set of $2N_\omega$ points $\{z_p\}$ on the complex plane with ${\rm Im}\,z_p >0$ consists of $N_\omega$ points on the imaginary axis at the fermionic Matsubara frequencies and $N_\omega$ equally spaced points on a circle with the radius $R=\pi(2N_\omega+5)/\beta$. This choice of data points alleviates the problems in the accuracy of the results and the stability of the convergence encountered in the above mentioned iteration process. The similar technique is also used in the analytic continuation with MEM in Sec.~\ref{MEM}. To find the optimal solution of the parameter set $\{\varepsilon^{\rm b}_l,V_l\}$ is not a straightforward task because of the presence of numerous solutions with nearly the same distance.  A genetic algorithm is applied in combination with the conjugate gradient method to improve the slow convergence of the solution. 

\section{Maximum Entropy Method\label{MEM}}
\begin{figure}
\includegraphics[width=8cm]{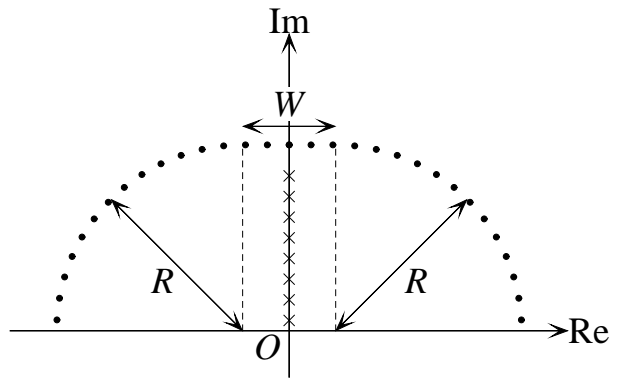}
\caption{\label{stadium}Schematic representation of the data points on the complex plane used for MEM. In addition to the data at the fermionic Matsubara frequencies $z_n=i(2n-1)\pi/\beta$ ($n=1,2,\cdots ,N_{\omega}$) depicted as the crosses on the imaginary axis, the equally spaced points represented as the dots on the curve consisting of two quarter circle arcs with the radius $R$ connected by the straight line with the length $W$ in Eq.\,(\ref{W}) are included in this study.}
\end{figure}
Since the perturbative calculations are performed with the Matsubara frequency in the present formalism for LDFA, the analytic continuation is required to convert the results as functions of the real frequency. For the spectral function $A(\omega)\equiv -(1/\pi){\rm Im}\,G(\omega)$, the relation to the Green's function at an arbitrary complex number $z$ 
\begin{align}
G(z)=\int_{-\infty}^{\infty} \frac{1}{z-\omega}A(\omega)\,d\omega
\end{align}
can be utilized.
 If a set of $N_G$ data of the Green's function $\bm{G}\equiv (G(z_1),G(z_2),\cdots)$ is given as input, one may obtain approximately a set of $N_A$ discretized data of spectral function $\bm{A}\equiv (A(\omega_1)\Delta\omega, A(\omega_2)\Delta\omega,\cdots)$ by solving the linear equation
\begin{align}
\bm{G}=K\bm{A}, \label{linear}
\end{align}
where $K$ denotes the $N_G\times N_A$ matrix with $K_{ij}\equiv 1/(z_i-\omega_j)$. Solving Eq.\,(\ref{linear}) is known to be severely ill-posed problem and the effective number of constraints imposed to $\bm{A}$ by the equation is far less than $N_G$ within the practical numerical precision. To extract the limited information about $\bm{A}$ properly, the maximum entropy method based on Bayesian inference is employed\cite{MJarrell1996}, where the entropic prior as a means of the regularization is introduced to circumvent the problem of the overfitting.

The joint probability of $\bm{A}$ and hyperparameters $\alpha_{\chi}$ and $\alpha_{S}$ for given $\bm{G}$ is described by using Bayes's theorem as 
\begin{align}
&P(\bm{A},\alpha_{\chi},\alpha_{S}|\bm{G})=\nonumber \\
&~~~~~~~~~~P(\bm{G}|\bm{A},\alpha_{\chi})P(\bm{A}|\alpha_{S})P(\alpha_{\chi})P(\alpha_S)/P(\bm{G}). \label{PAG}
\end{align}
The distribution of the sum of squared relative errors 
\begin{align}
\chi^2\equiv \sum_{i=1}^{N_G}\frac{\left|G_i-\sum_j K_{ij}A_j\right|^2}{|G_i|^2} \label{chi2}
\end{align}
is assumed to be represented as a Gaussian function 
\begin{align}
P(\bm{G}|\bm{A},\alpha_{\chi})=\frac{\alpha_{\chi}^{N_G/2}}{(2\pi)^{N_G/2}\prod_i|G_i|}\exp(-\alpha_{\chi}\chi^2/2). \label{PGA}
\end{align}
Here,  $\alpha_{\chi}$ is also optimized as a hyperparameter, since the standard deviation $\sigma=1/\sqrt{\alpha_{\chi}}$ for the spectral function $A_{\bm{k}}(\omega)$ inferred by MEM from the LDFA Green's function of the 2D Hubbard model has strong momentum $\bm{k}$ dependence: while the values of $\sigma$ at $\bm{k}$ points on the Fermi surface, i.e., the X--M$_2$ line in Fig.~\ref{AK}(d), range from $\sigma=5\times 10^{-7}$ to $2\times 10^{-6}$,  those at $\bm{k}=(0,0)$ are from  $\sigma=3\times 10^{-5}$ to $4\times 10^{-4}$.
The entropic prior is given as
\begin{align}
P(\bm{A}|\alpha_{S})\approx \frac{\alpha_S^{N_A/2}}{(2\pi)^{N_A/2}\prod_i m_i^{1/2}} \exp(\alpha_{S}S),  \label{PS}
\end{align}
where $S$ is the relative entropy between $\bm{A}$ and the default model $\bm{m}\equiv(m_1,m_2,\dots)$, and is defined as
\begin{align}
S=\sum_{i=1}^{N_A} \left[A_i - m_i-A_i\ln(A_i/m_i)\right].\label{entropy}
\end{align}
The uniform distribution is adopted for the default model: $m_i=1/N_A$. The prior probabilities for hyperparameters $P(\alpha_{\chi})$ and $P(\alpha_S)$ are assumed to be constants and $P(\bm{G})$ is the normalization factor. 

The joint probability of the hyperparameters $\alpha_{\chi}$ and $\alpha_{S}$ can be obtained inserting Eqs.\,(\ref{chi2})--(\ref{entropy}) into Eq.\,(\ref{PAG}) and integrated it over $\bm{A}$ within the Gaussian approximation:
\begin{align}
&\ln P(\alpha_{\chi},\alpha_{S}|\bm{G}) \approx\textrm{const.} +\frac{N_G}{2}\ln\alpha_{\chi}\nonumber \\
&~~~~+\frac{1}{2}\sum_{i=1}^{N_A}\ln\frac{\alpha_SA^{\textrm{max}}_i}{\alpha_{\chi}\lambda_i+\alpha_S} -\frac{\alpha_{\chi}}{2}\chi^2( \bm{A}^{\textrm{max}})+\alpha_SS(\bm{A}^{\textrm{max}}),\label{PchiS}
\end{align}
where $\bm{A}^{\textrm{max}}$ denotes $\bm{A}$ at which $P(\bm{A},\alpha_{\chi},\alpha_{S}|\bm{G})$ is the maximum with given values of $\alpha_{\chi}$ and $\alpha_{S}$. $\lambda_i$ represents  $i$th eigenvalue of the matrix $\Lambda$:
\begin{align}
\Lambda_{ij}=\sum_{l=1}^{N_G} (A^{\textrm{max}}_i)^{1/2}K_{li}^{*}K_{lj}(A^{\textrm{max}}_j)^{1/2}/|G_l|^2.
\end{align}

Our problem of the analytic continuation is now reduced to find the set of $\bm{A}$, $\alpha_{\chi}$ and $\alpha_{S}$ which is at the maximum of $P(\bm{A},\alpha_{\chi},\alpha_{S}|\bm{G})$. To do so, we first set guess values of $\alpha_{\chi}$ and $\alpha_{S}$ and maximize $P(\bm{A},\alpha_{\chi},\alpha_{S}|\bm{G})$ with respect to $\bm{A}$. This is equivalent to minimizing $Q(\bm{A})\equiv\alpha_{\chi}\chi^2(\bm{A})/2-\alpha_SS(\bm{A})$, which can be achieved using the Newton-Raphson method. The calculations are repeated with different values of $\alpha_{S}$ to find the maximum of $P(\alpha_{\chi},\alpha_{S}|\bm{G})$ in Eq.\,(\ref{PchiS}) with respect to $\alpha_{S}$ with fixed value of $\alpha_{\chi}$. The golden section search method is applied for this optimization of $\alpha_{S}$.
To find the maximum of $P(\alpha_{\chi},\alpha_{S}|\bm{G})$ in Eq.\,(\ref{PchiS}) with respect to $\alpha_{\chi}$, 
\begin{align}
\frac{\partial}{\partial\alpha_{\chi}}P(\alpha_{\chi},\alpha_{S}|\bm{G})=0
\end{align}
is calculated assuming the $\alpha_{\chi}$ dependence of $\bm{A}^{\textrm{max}}$ is negligible. The resultant equation for the optimal $\alpha_{\chi}$ is
\begin{align}
\alpha_{\chi}=\frac{1}{\chi^2( \bm{A}^{\textrm{max}})}\left({N_G}-\sum_{i=1}^{N_A}\frac{\lambda_i}{\lambda_i+\alpha_S/\alpha_{\chi}}\right). \label{alfchi}
\end{align}
This can be solved iteratively by inserting previously obtained value of $\alpha_{\chi}$ repeatedly on the right-hand side of the equation. After solving Eq.\,(\ref{alfchi}), the optimization of $\bm{A}$ and $\alpha_{S}$ follows with this new value of $\alpha_{\chi}$ and again solving Eq.\,(\ref{alfchi}). This process is repeated until convergence is reached.

In the LDFA calculations, the $N_{\omega}$ data of $G(z_n)$ at the fermionic Matsubara frequencies, i.e. $z_n=i(2n-1)\pi/\beta$ ($n=1,2,\cdots,N_{\omega}$) on the imaginary axis are adopted.  For the rest of the data, instead of taking them on the Matsubara frequencies, a set of data points placed at the same distance $R$ from the nearest pole of $G(z)$ on the real axis is chosen: equally spaced points on the curve consisting of two quarter circle arcs with the radius $R$ connected by a straight line with the length $W$ in Eq.\,(\ref{W}) as shown in Fig.\,\ref{stadium}. This choice of data points ameliorates the difficulty of solving Eq.\,(\ref{linear}) and the accuracy of $A(\omega)$ obtained is improved, particularly in structures away from the Fermi level. The value of $R$ adopted in this study is $R=2.7W$ except for $U=3.0$, for which $R=1.8W$ is used and the number of these additional points is $N_{\rm R}\approx R\beta$.

Note that altering the default model other than the uniform distribution scarcely affects the results. What is even more important is the model for $\chi^2$ in Eq.\,(\ref{chi2}), where the square sum of the relative errors is assumed instead of the absolute errors as in the previous studies and the data points are chosen as in Fig.~\ref{stadium}. This is probably due to the different quality of the data we obtained here, where the main source of errors comes from arithmetic operations and statistical errors are absence unlike previous studies with QMC\cite{MJarrell1996}.
The DOS of the 2D Hubbard model inferred by the present and standard  maximum entropy methods are compared in Appendix D.

\section{2D Hubbard Model\label{2DHubbard}}
\subsection{Accuracy of the results and bench mark\label{ssBench}}

Before discussing the LDFA results of the 2D Hubbard model obtained with the new Lanczos ED method in detail, here, we evaluate the accuracy of the results and make comparisons with results of other methods. We first check the accuracy of the four-point vertex function in Eq.\,(\ref{vtx4def}). Since we are dealing with the PM state, the spin rotational symmetry can be verified. To do this, the difference between the horizontal-spin $\gamma^{\uparrow\uparrow\uparrow\uparrow}-\gamma^{\uparrow\downarrow\downarrow\uparrow}$ and vertical-spin $\gamma^{\uparrow\downarrow\uparrow\downarrow}$ components, which should be zero for the exact calculations, are examined with respect to $\Omega$ as
\begin{align}
\varepsilon(\Omega)=\frac{\sum_{\omega,\omega'}\big|\gamma^{\uparrow\uparrow\uparrow\uparrow}_{\omega\omega';\Omega}-\gamma^{\uparrow\downarrow\downarrow\uparrow}_{\omega\omega';\Omega}-\gamma^{\uparrow\downarrow\uparrow\downarrow}_{\omega\omega';\Omega}\big|}{\sum_{\omega,\omega'}\left\{
\big|\gamma^{\uparrow\uparrow\uparrow\uparrow}_{\omega\omega';\Omega}\big|
+\big|\gamma^{\uparrow\downarrow\downarrow\uparrow}_{\omega\omega';\Omega}\big|
+\big|\gamma^{\uparrow\downarrow\uparrow\downarrow}_{\omega\omega';\Omega}\big|\right\}}.
\end{align}
It is found that $\varepsilon(\Omega)\sim 10^{-5}$ for the lowest $\Omega\sim 0$ and the highest $\Omega=4N_{\omega}\sim 4W$ frequencies and less accurate $\varepsilon(\Omega)\sim 10^{-4}$ for intermediate frequencies $\Omega \sim 2W$. This is expected from the approximation made in the  Lanczos ED method, which is accurate in low-energy excitations and asymptotic behavior as mentioned in Sec.~\ref{LanczosED}.  

Similarly, the accuracy of the LDFA results can be assessed by calculating the same quantity with different directions of spin for the PM state. The accuracy is  typically about 6 digits for values such as the double occupancy $D\equiv\langle n_{i\uparrow}n_{i\downarrow}\rangle$ calculated using the Migdal-Galitskii formula\cite{Galitskii1958}.
At low temperatures, however, because of the divergent property of the spin susceptibility $\chi_{\rm sp}\sim e^{\Delta/T}$ as $T\rightarrow 0$\cite{SChakravarty1988,PHasenfratz1991},  the Bethe-Salpeter equation in Eq.\,(\ref{BSE}) is unstable when $\chi_{\rm sp}^{-1}\propto 1-\lambda_{\rm sp}\approx e^{-\Delta/T}$ is too small  $1-\lambda_{\rm sp}\lesssim 10^{-3}$, where $\lambda_{\rm sp}$ is the maximum eigenvalue of a $2N_{\omega} \times 2N_{\omega}$ matrix
\begin{align}
M_{\omega,\omega'}\equiv\frac{1}{\beta}\gamma^{({\rm sp})}_{\omega,\omega';0}\chi^{d,0}_{(\pi,\pi),\omega',0}. 
\end{align}
Although the problem can be partly avoided by using technique in Ref.~\onlinecite{JOtsuki2014}, obtained results are less accurate.

As mention before, in the ED method the conduction band is replaced by a finite number of the bath energy levels in the effective IAM.  While the accuracy of the results is expected to increase as the number of the bath levels $N_{\rm b}$ increases, numerical errors introduce by the Lanczos algorithm, where high-energy excitations are omitted, would increase as the number of basis function increases. To check the convergence of the results as a function of $N_{\rm b}$, the values of $D$ obtained with $N_{\rm b}=3$, 5, and 7 have been compared.  Note that one of the bath level is required to be placed at the Fermi level to describe a metallic state and thus $N_{\rm b}$ needs to be an odd number for the half-filled square-lattice Hubbard model because of the electron-hole symmetry\cite{HHafermann2010}. Whereas considerable difference $|D_{N_{\rm b}=3}-D_{N_{\rm b}=5}|/D_{N_{\rm b}=5} \sim 10^{-3}$ between those obtained with $N_{\rm b}=3$ and $N_{\rm b}=5$ is found for $U=4.0$ at low temperatures, the discrepancy between those obtained with $N_{\rm b}=5$ and $N_{\rm b}=7$ is already within $2\times 10^{-5}$. 

The deviations from the hybridization sum rule\cite{E.Koch2008}, which relates the hybridization strength of the effective IAM and the lattice hopping integrals, have also been examined. The sum rule for the square-lattice Hubbard model with the nearest-neighbor hopping is 
\begin{align}
\sum_{l=1}^{N_{\rm b}} V_l^2=4t^2.
\end{align}
The deviation is rapidly reduced with increasing $N_{\rm b}$: the relative errors $\varepsilon=|(\sum_l V_l^2)^{1/2}-2t|/2t$ are utmost $\varepsilon=0.15$, $6\times 10^{-3}$, and $6\times 10^{-5}$ for $N_{\rm b}=3$, 5, and 7, respectively. These findings indicate that well converged values can be obtained with $N_{\rm b}=7$. Hence, all results presented in the rest of the paper were calculated with $N_{\rm b}=7$. The number of discretized momentum points in the Brillouin zone of the square lattice used in the calculations is 64$\times$64. For simplicity, we set the value of the hopping integral $t=1$.

\begin{figure}
\includegraphics[width=8cm]{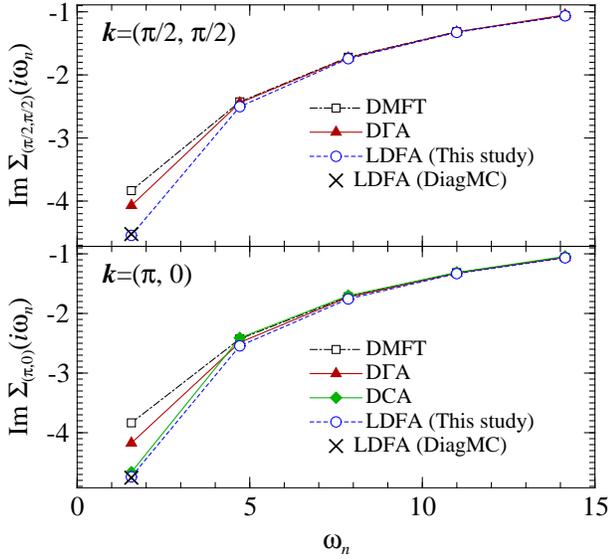}
\caption{\label{g1kcmp}Comparison of  the imaginary part of the self-energies ${\rm Im}\,\Sigma_{\bm{k}}(i\omega_n)$ on the imaginary axis at momenta $\bm{k}=(\pi/2,\pi/2)$ (upper panel) and $\bm{k}=(\pi,0)$ (lower panel) for $U=8.0$ and $\beta=2.0$ obtained with various methods. Those of LDFA (open circles) and DMFT (open squares) are the results of this study. The DCA data (closed diamonds) are reproduced from Ref.~\onlinecite{JPFLeBlanc2015}, and the D$\Gamma$A data (closed triangles) from Ref.~\onlinecite{TSchaefer2016}. The LDFA with the diagrammatic QMC results  (crosses) are taken from Ref.~\onlinecite{JGukelberger2017} (only data with the Matsubara frequency of $\omega_1=\pi/\beta$ are available).}
\end{figure}
In Fig.~\ref{g1kcmp}, the imaginary part of the self-energy ${\rm Im}\,\Sigma_{\bm{k}}(i\omega_n)$ on the imaginary axis at momenta $\bm{k}=(\pi/2,\pi/2)$ and $\bm{k}=(\pi,0)$ for $U=8.0$ and $\beta=2.0$ obtained with various methods are compared. The differences are mainly found at the lowest Matsubara frequency, i.e., $\omega_1\equiv\pi/\beta$. The LDFA values of ${\rm Im}\,\Sigma_{\bm{k}}(i\omega_n)$ are substantially reduced from that of DMFT at $\omega_1$ and are in good agreements with the DCA result for $\bm{k}=(\pi,0)$ in Ref.~\onlinecite{JPFLeBlanc2015}. Our LDFA results are also in good agreement with the previous LDFA results with the diagrammatic QMC method in Ref.~\onlinecite{TSchaefer2016}. On the other hand, the D$\Gamma$A values  are placed between those of DMFT and LDFA. The D$\Gamma$A values at $\omega_1$ considerably deviate from that of DCA.

\subsection{Energetics\label{ssD}}

It has been a long-standing debate over where and how the crossover or transition from the Slater to Mott-Heisenberg regime occurs as $U$ increases in the 2$D$ Hubbard model\cite{PWAnderson1997,KBorejsza2003,LFratino2017,TSchaefer2015,EGCPvanLoon2018,SMoukouri2001,BKyung2003,EGull2008,LFTocchio2016}. In the Slater regime electrons are delocalized at high temperatures. In the weak coupling limit, the potential energy decreases in the presence of AFM correlations and therefore the stabilization of AFM order is expected to be mainly driven by the potential energy in the Slater regime. On the other hand, in the Mott-Heisenberg regime the localization of electrons already occurs at high temperatures and local spins are formed. In the strong coupling limit, the AFM coupling between the localized spins can be regarded as a virtual process through the electron hopping and the stabilization of AFM order is, therefore, expected to be mainly driven by the kinetic energy in the Mott-Heisenberg regime. In this subsection, we discuss the temperature and $U$ dependence of the double occupancy and the kinetic energy obtained with LDFA.

\begin{figure}
\includegraphics[width=7.5cm]{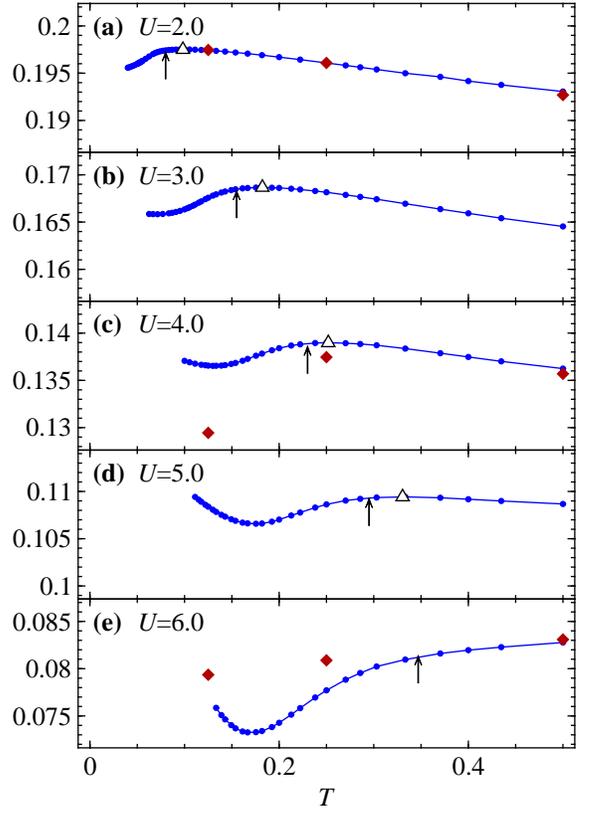}
\caption{\label{DT}Temperature dependence of the double occupancy $D$ for $U=2.0$ (a), $U=3.0$ (b), $U=4.0$ (c), $U=5.0$ (d) and $U=6.0$ (e). The local maxima are shown by the triangles in (a)--(d). In each panel, the arrow indicates $T^{\rm DMFT}_{\rm N}$ taken from Fig.~1 in Ref.~\onlinecite{JOtsuki2014}.  For comparison, the values of $D$ calculated with DCA in Ref.~\onlinecite{JPFLeBlanc2015} are also represented by the diamonds in (a), (c), and (e).}
\end{figure}
Figure~\ref{DT} shows temperature dependence of the double occupancy $D\equiv\langle n_{i\uparrow}n_{i\downarrow}\rangle$ for various values of $U$ calculated using the Migdal-Galitskii formula\cite{Galitskii1958}. For $U\le 5.0$ at high temperature, $D$ increases with decreasing temperature, reaches its local maximum, and then decreases. The local maximum is positioned just above the DMFT N{\'e}el temperature $T^{\rm DMFT}_{\rm N}$ indicated by the vertical arrows. Although no long-range AFM order is presence in LDFA in finite temperatures, which abides by the Mermin-Wagner theorem, the connection between the local maximum of $D$ found at temperature near $T^{\rm DMFT}_{\rm N}$ for $U\le 5.0$ and AFM correlations is apparent. The presence of the local maximum is also found in DCA study for $U=4.0$ in Ref.~\onlinecite{JPFLeBlanc2015} as indicated by the diamonds in Fig.\,\ref{DT}(c). 
The temperature dependence of $D$ for $U\le 5.0$ is consistent with what expected in the Slater regime; in high-temperature metallic state, because of the Fermi degeneracy, $D$ increases with decreasing temperature and $D$ decreases as AFM correlations develop below $T^{\rm DMFT}_{\rm N}$, where electrons efficiently avoid in each other. These findings are also consistent with the previous LDFA study, where the reduction of the potential energy due to nonlocal AFM correlations found in a range $1.0 \le U\le 4.0$\cite{EGCPvanLoon2018}.

\begin{figure}
\includegraphics[width=8cm]{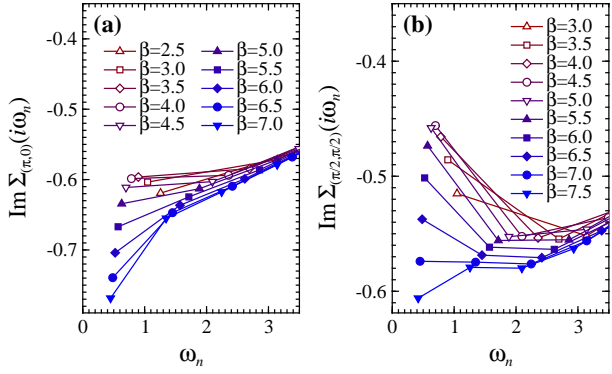}
\caption{\label{SGMkU40}The imaginary part of the self-energy ${\rm Im}\,\Sigma_{\bm{k}}(i\omega_n)$ for $U=4.0$ with various values of $\beta$ at momenta on the Fermi surface $\bm{k}=(\pi,0)$ (the X point) (a) and $\bm{k}=(\pi/2,\pi/2)$ (the M$_2$ point) (b). }
\end{figure}
To elucidate the relation between the AFM correlation and the temperature dependence of $D$ more in detail, in Fig.\,\ref{SGMkU40}, the imaginary part of the self-energy ${\rm Im}\,\Sigma_{\bm{k}}(i\omega_n)$ for $U=4.0$ at momenta on the Fermi surface $\bm{k}=(\pi,0)$ (the X point) and $\bm{k}=(\pi/2,\pi/2)$ (the M$_2$ point) are shown for various values of $\beta$ in panels (a) and (b), respectively. For a non-correlated metal, ${\rm Im}\,\Sigma_{\bm{k}}(i\omega_n)$ at the lowest Matsubara frequency $\omega_1=\pi/\beta$ is expected to increase as temperature is lowered, because of the reduction of thermal fluctuations. Hence, the reduction of ${\rm Im}\,\Sigma_{\bm{k}}(i\pi/\beta)$ found in Fig.\,\ref{SGMkU40} indicates the increase of AFM fluctuations and the temperature where ${\rm Im}\,\Sigma_{\bm{k}}(i\pi/\beta)$ takes the local maximum would be regarded as the onset temperature of AFM correlations. Indeed, the local maximum ${\rm Im}\,\Sigma_{\bm{k}}(i\pi/\beta)$ temperature of the M$_2$ point coincides with $T_{\rm N}^{\rm DMFT}$ within the deviation of 0.01 in the range of $U=$2.0 to 6.0. It is also found in the previous LDFA study that the maximum of the uniform spin susceptibility for $U=2.0$, 3.0 and 4.0 is located at temperatures close to $T_{\rm N}^{\rm DMFT}$\cite{EGCPvanLoon2018}. On the other hand, the local maximum temperature of ${\rm Im}\,\Sigma_{\bm{k}}(i\pi/\beta)$ of the X point is placed higher than $T_{\rm N}^{\rm DMFT}$ and roughly follows the local maximum temperature of $D$.

In contrast, for $U=6.0$, no local maximum can be found in $D$ below $T=0.5$ in Fig.\,\ref{DT}(e): $D$ monotonically decreases around $T^{\rm DMFT}_{\rm N}$ with decreasing temperature and increases at much lower temperature ($T<0.17$). As can be seen Figs.\,\ref{DT}(d) and \ref{DT}(e), the increase of $D$ with decreasing temperature occurs much lower temperature than $T^{\rm DMFT}_{\rm N}$ for $U\ge 5.0$ and the relation between the onset of the short-range AFM order and the temperature dependence of $D$ is less clear for $U\ge 5.0$.
The lack of local maximum of $D$ for $U\ge 5.5$ below $T=0.5$ coincides with the absence of the local maximum  ${\rm Im}\,\Sigma_{\bm{k}}(i\pi/\beta)$ of the X point and indicates non-Fermi-liquid or ``bad metallic'' behavior at high temperatures. The fact is consistent with the Mott-Heisenberg regime, where local spins are expected to be preformed above the onset temperature of AFM correlations, i.e., $T_{\rm N}^{\rm DMFT}$.

\begin{figure}
\includegraphics[width=8cm]{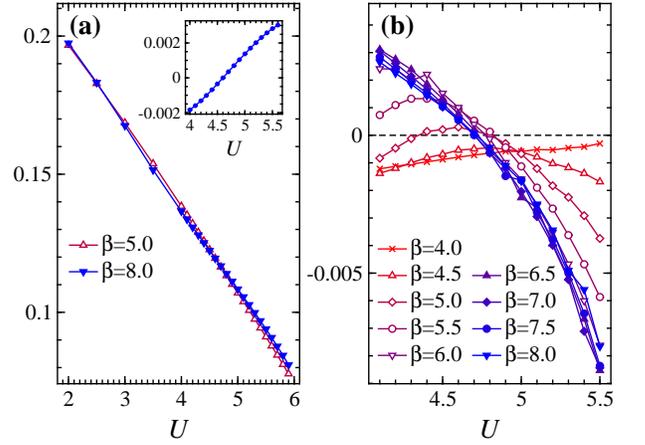}
\caption{\label{DU}(a) Comparison between the $U$ dependence of the double occupancy $D$ at $\beta =5.0$ (open tip-up triangles) and $\beta =8.0$ (closed tip-down triangles); The inset shows the values of $D$ at $\beta =5.0$ subtracted from that of $\beta =8.0$ $D_{\beta =8.0}-D_{\beta =5.0}$ in a range from $U=4.0$ to $U=5.6$. (b) $\partial^2 D/\partial U^2$ as a function of $U$ for various values of $\beta$.}
\end{figure}
To clarify where the crossover from the Slater to Mott-Heisenberg regime occurs, in Fig.~\ref{DU}(a), $D$ as a function of $U$ are plotted for $\beta =5.0$ and $\beta =8.0$. As can be seen in the $D$ curve of $\beta =5.0$, $D$ decreases linearly with increasing $U$ at high temperatures.  However, at low temperatures, the decrease of $D$ does not evenly happen as can be observed in the $D$ curve of $\beta =8.0$: the faster (slower) reduction of $D$ for $U<4.7$  ($U>4.7$) causes the formation of the concave (convex) in the $D$ curve. This variation of $D$ at low temperatures is in accordance with the crossover from the Slater to Mott-Heisenberg regime with increasing $U$ and thus one can regard the inflection point of the $D$ curve as a boundary between the two regimes. To examine the inflection point of $D$ curve more in detail,  in Fig.~\ref{DU}(b), $\partial^2 D/\partial U^2$ as a function of $U$ is shown for various values of $\beta$. For $\beta =4.0$, there is no inflection point in the $D$ curve in a range from $U=4.1$ to $U=5.5$. As temperature decreases, however, $\partial^2 D/\partial U^2$ as a function of $U$  rapidly converges into a curve, which intersects the $\partial^2 D/\partial U^2$=0 line at $U\approx 4.7$, and changes its sign from positive to negative at $U\approx 4.7$. These results indicates that the crossover from the Slater to Mott-Heisenberg regime occurs around $U^{*}\approx 4.7$.
Similar crossover from the Slater to Mott-Heisenberg regime has been also found in the AFM phase in the CDMFT\cite{LFratino2017} and variational QMC\cite{LFTocchio2016} studies.

\begin{figure}
\includegraphics[width=8cm]{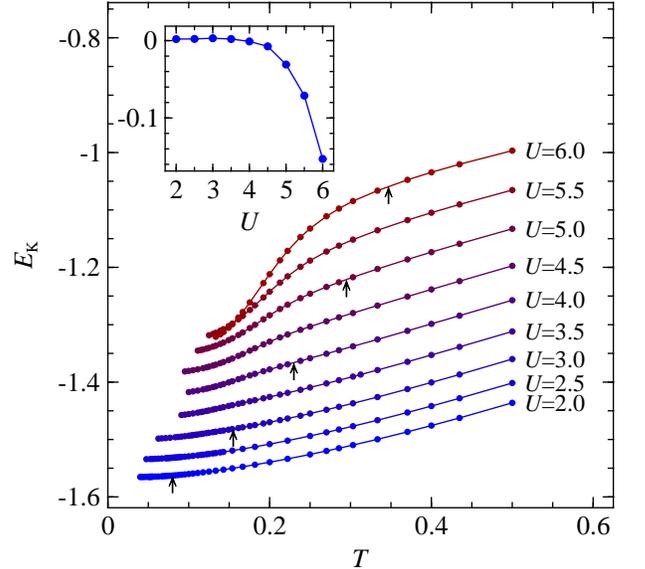}
\caption{\label{EK}Kinetic energy $E_{\rm K}$ as functions of temperature $T$ for various values of $U$. The arrows indicate $T^{\rm DMFT}_{\rm N}$ taken from Fig.~1 in Ref.~\onlinecite{JOtsuki2014}. The inset shows $\Delta E_{\rm K}$ for various $U$.}
\end{figure}
Figure~\ref{EK} shows temperature dependence of the kinetic energy $E_{\rm K}$ for various values of $U$. For small $U$, the inclination of $E_{\rm K}$ is reduced with decreasing temperature and no clear change of this tendency at $T^{\rm DMFT}_{\rm N}$ is found. In contrast, for $U>4.5$ the inclination of $E_{\rm K}$ increases with decreasing temperature particularly for $T<T^{\rm DMFT}_{\rm N}$ resulting in steep precipitation of $E_{\rm K}$ below $T^{\rm DMFT}_{\rm N}$. This tendency becomes more clear as $U$ increases.
To make a rough estimation of the lowering of $E_{\rm K}$ caused by the AFM correlation, we assume $E_{\rm K}$ without the AFM correlation can be approximated by a linear function of $T$ below $T^{\rm DMFT}_{\rm N}$ and subtract this approximated value from $E_{\rm K}$ at the lowest temperature $T_{\rm L}$ available as
\begin{align}
\Delta E_{\rm K}&\equiv  E_{\rm K}(T_{\rm L})-E_{\rm K}(T_{\rm N}^{\rm DMFT})\nonumber\\
&~~~~~-\frac{\partial E_{\rm K}}{\partial T}\Bigg|_{T=T_{\rm N}^{\rm DMFT}}(T_{\rm L}-T_{\rm N}^{\rm DMFT}).
\end{align}
The result is presented in the inset of Fig.~\ref{EK}.  Whereas $\Delta E_{\rm K}$ remains small $|\Delta E_{\rm K}| <10^{-2}$ for $U\le 4.5$, $\Delta E_{\rm K}$ rapidly decreases with increasing $U$ for $U>5.0$. This fact is consistent with the crossover behavior of $D$ from the Slater to Mott-Heisenberg regime around $U^{*} \approx 4.7$, since in the latter the stabilization of the short-range AFM order is expected to be driven by the kinetic energy.

\subsection{Structures of DOS and the spectral function\label{ssAk}}
\begin{figure*}
\includegraphics[width=4cm]{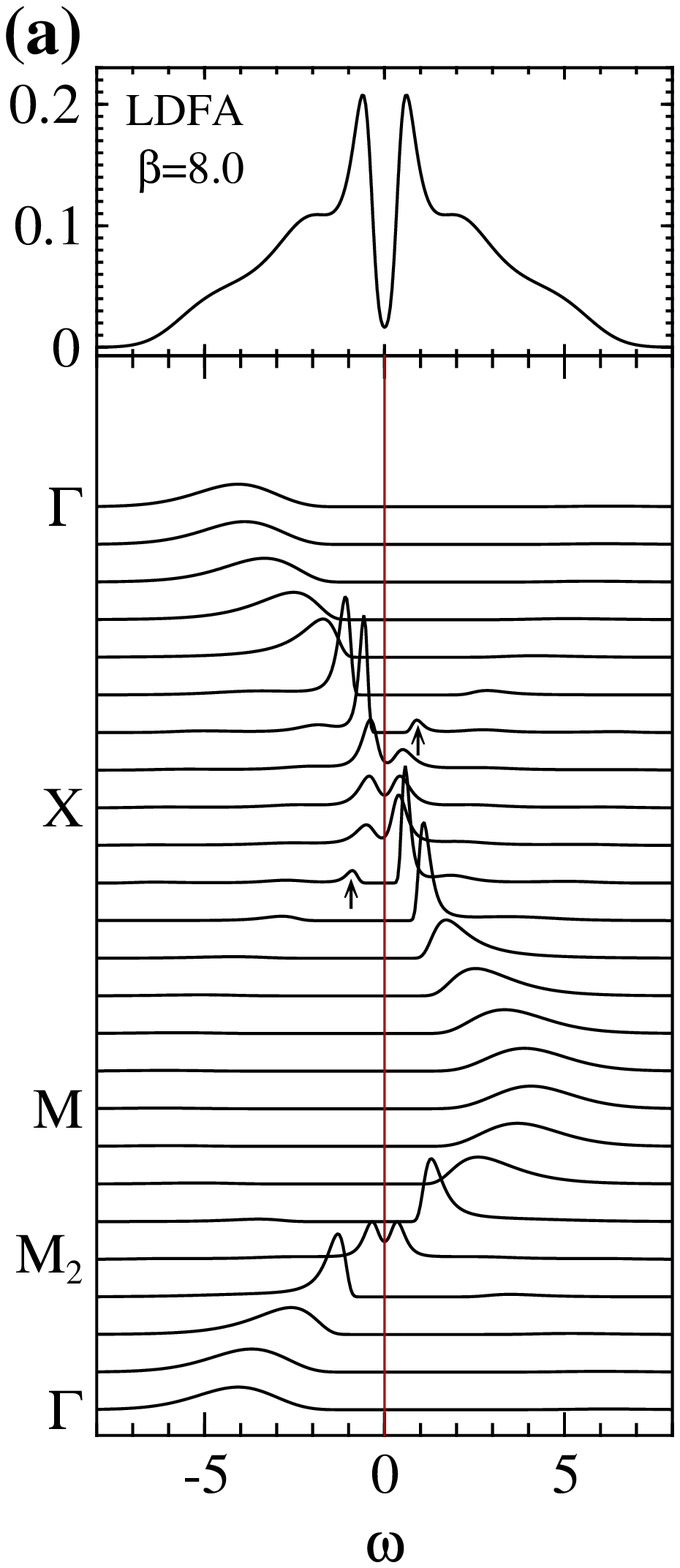}~~
\includegraphics[width=4cm]{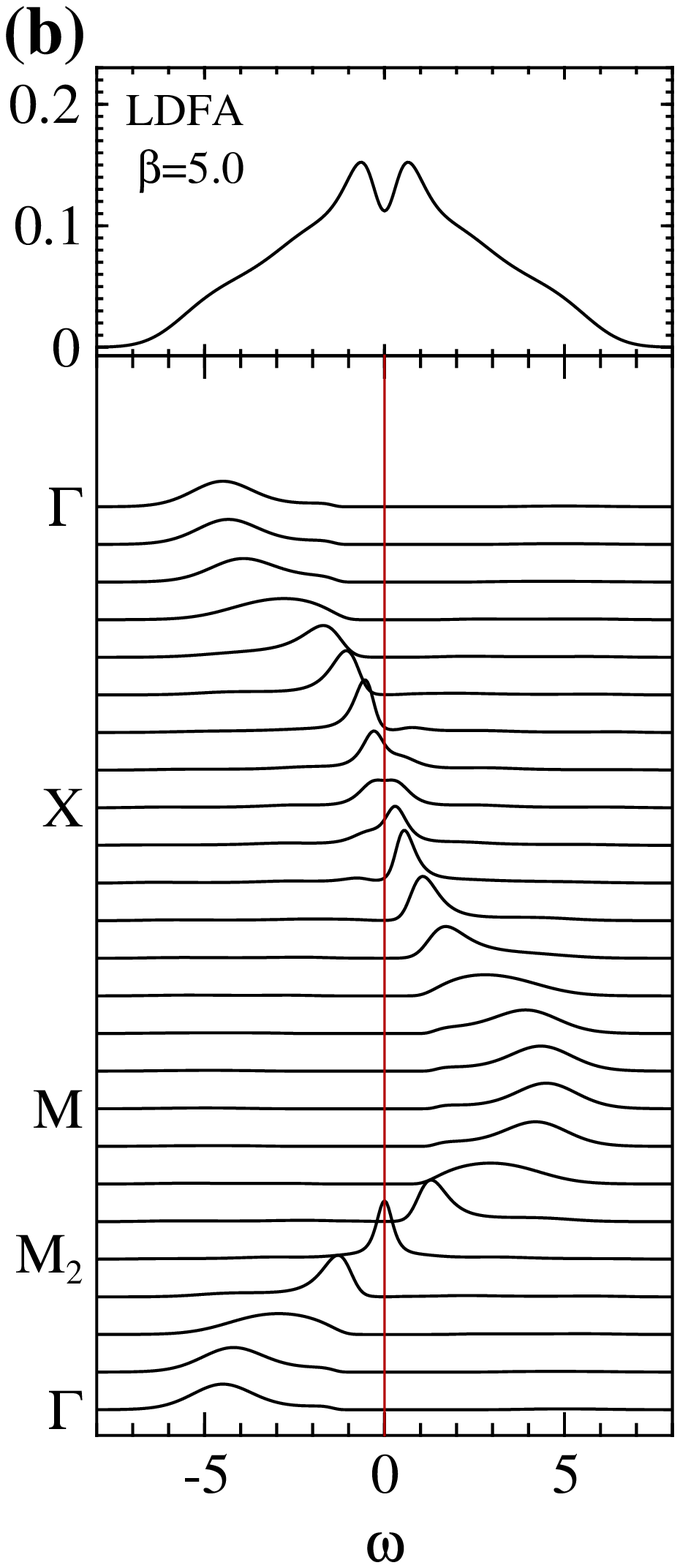}~~
\includegraphics[width=4cm]{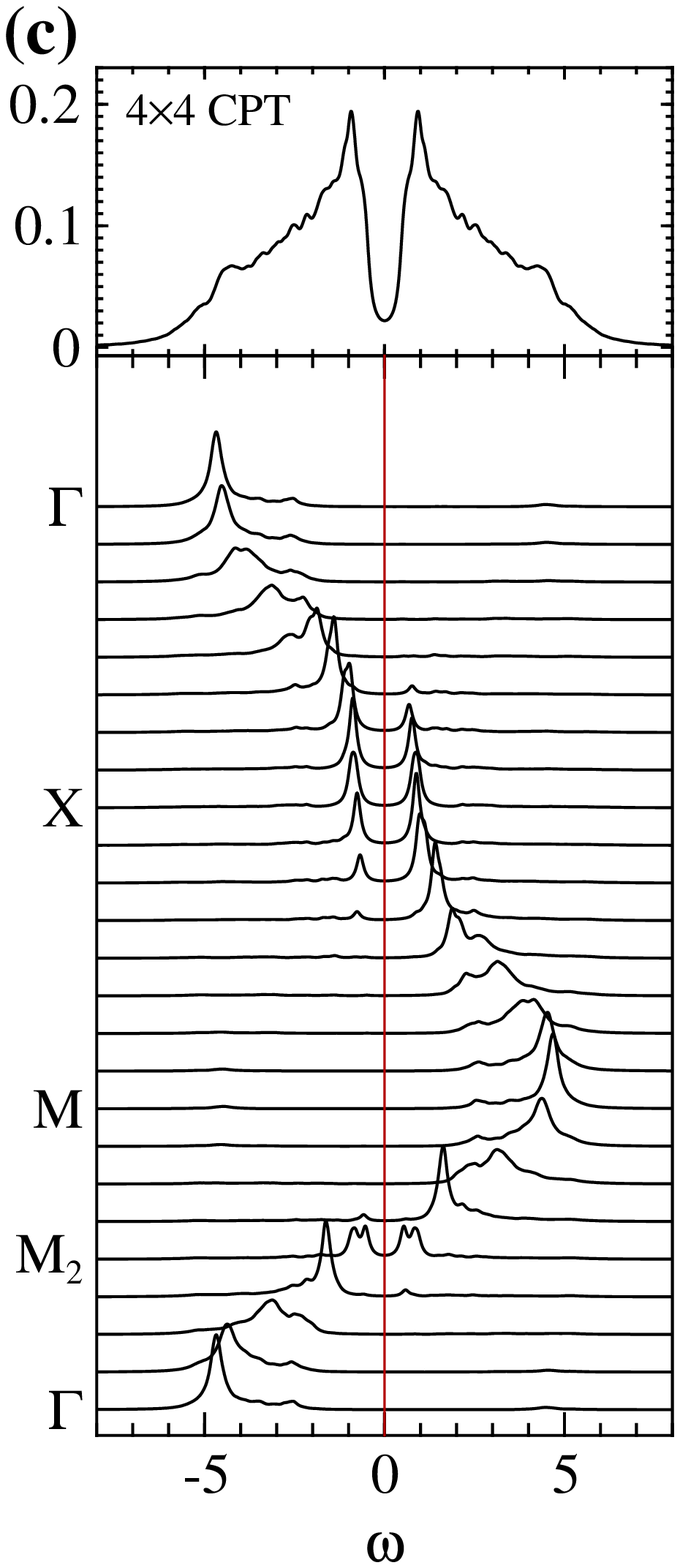}~~
\includegraphics[width=3.4cm]{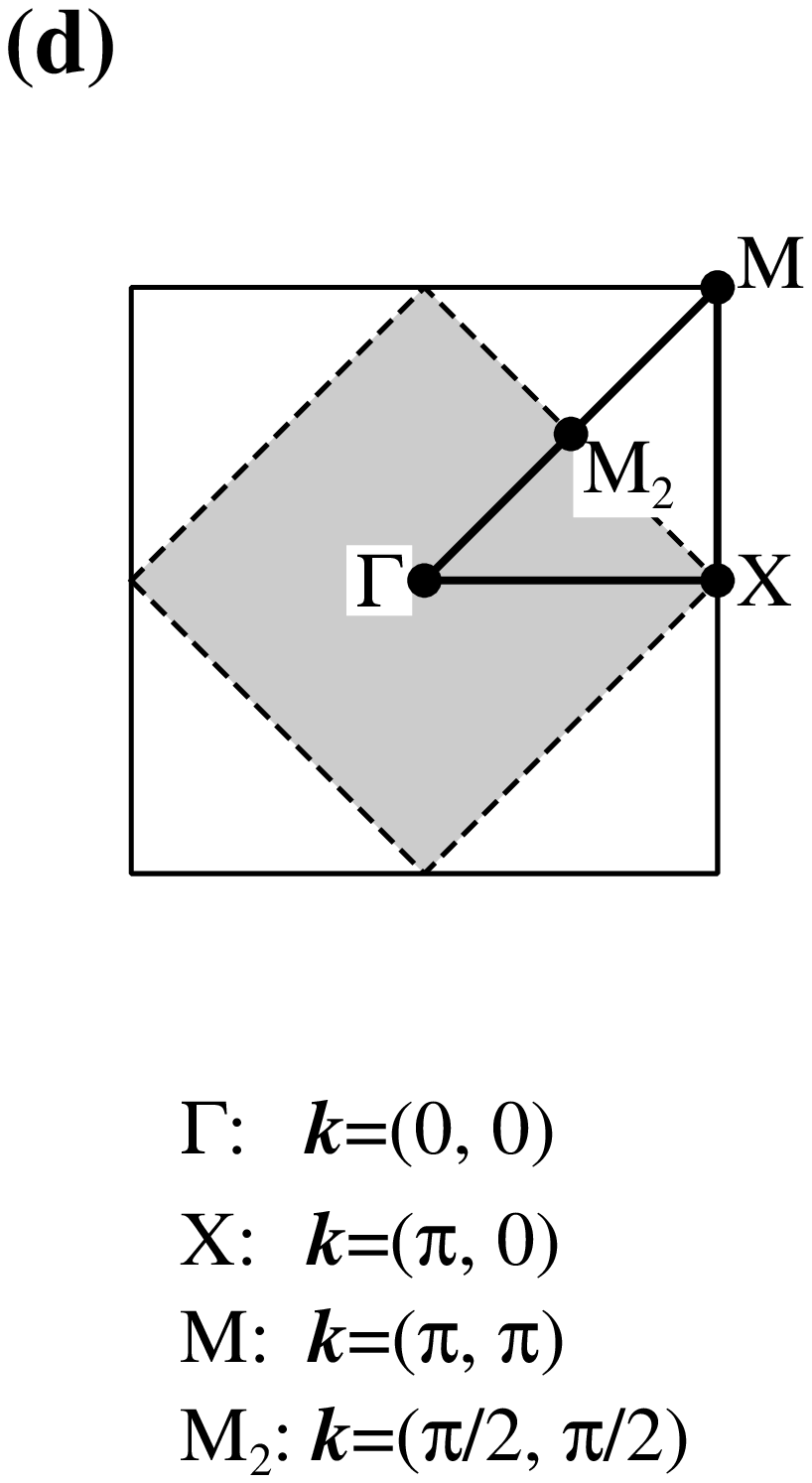}
\caption{\label{AK}DOS $\rho(\omega)$ (top) and the spectral functions $A_{\bm{k}}(\omega)$  with momenta $\bm{k}$ along the symmetry lines $\Gamma$--X, X--M and M--$\Gamma$ (bottom) for $U=4.0$ calculated by means of LDFA with $\beta=8.0$ (a), LDFA with $\beta=5.0$ (b) and CPT with a $4\times 4$ cluster (c). In (d), the Brillouin zone of the square lattice and the symmetry lines and points are depicted; the dashed line represents the Fermi surface at half filling for $U=0$.}
\end{figure*}
Figure~\ref{AK} shows LDFA results of DOSs and the spectral functions $A_{\bm{k}}(\omega)$ with momenta $\bm{k}$ along the symmetry lines $\Gamma$--X, X--M and M--$\Gamma$ for $U=4.0$ with $\beta=8.0$ in panel (a) and with $\beta=5.0$ in panel (b). For comparison those obtained with the cluster perturbation theory (CPT)\cite{DSenechal2000,DSenechal2002} with a $4\times 4$ cluster are also presented in panel (c).
Although the pseudogap already exists in DOS with $\beta=5.0$, which is also found in the previous LDFA studies\cite{HHafermann2010,EGCPvanLoon2018}, one can see the prominent development of the pseudogap DOS for $\beta=8.0$. Such drastic development in the pseudogap with decreasing temperature is consistent with exponential growth of the AFM correlation length with decreasing temperature discussed in the D$\Gamma$A\cite{TSchaefer2015} and LDFA\cite{TRibic2018} studies. The formation of the pseudogap in DOS is found to start at the temperature close to $T_{\rm N}^{\rm DMFT}$. 

The formation of the pseudogap is also found in the quasiparticle peaks at the Fermi level in $A_{\bm{k}}(\omega)$ with $\bm{k}=(\pi,0)$ (the X point) and $\bm{k}=(\pi/2,\pi/2)$ (the M$_2$ point) for $\beta=8.0$. The pseudogap opening along the $\Gamma$-X and X-M lines in the vicinity of the X point is also seen along with the peaks corresponding to the shadow band (indicated by the arrows), which is the reminiscence of the Brillouin zone folding caused by the long-range AFM order. On the other hand, for $\beta=5.0$, although the incipience of pseudogap formation, e.g., the flattening of the quasiparticle peak top, can be observed at the X point, still no such indication found in the peak at the M$_2$ point. The formation of the pseudogap occurs first at the X point, spreads through the Fermi surface and ends at the M$_2$ point with decreasing temperature. The same trends in the temperature and momentum dependence in the pseudogap formation is found in the previous works with TPSC\cite{SMoukouri2000}, D$\Gamma$A\cite{AAKatanin2009} and the QMC calculations of finite-size clusters\cite{DRost2012}. More details of the momentum dependence of the pseudogap formation will be discussed in Sec.~\ref{ssPG}. 
Although  the structures of the LDFA spectral functions of the momenta near the $\Gamma$ point are blurred because of inaccuracy of the data and the size of the pseudogap in the vicinity of the Fermi level is substantially smaller compared to those of CPT, reasonable agreements can be found between those obtained with LDFA in panel (a) and CPT in panel (c). 

\subsection{Gap formation in DOS\label{ssGap}}
\begin{figure}
\includegraphics[width=8cm]{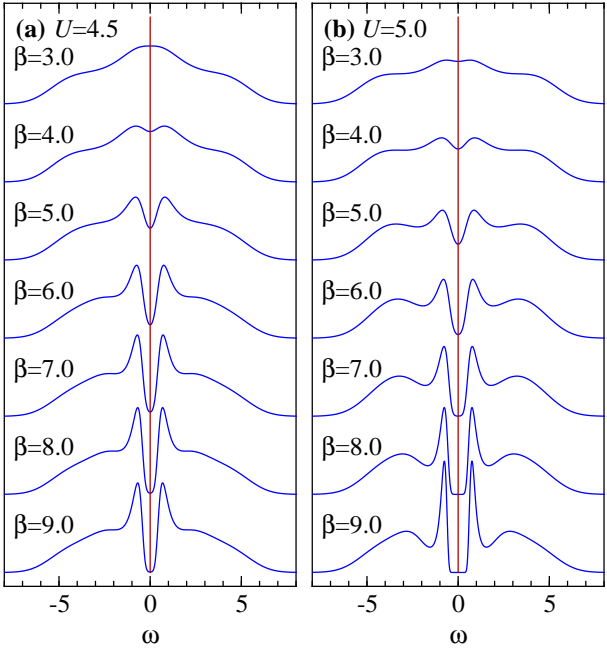}
\caption{\label{DOS}$\rho(\omega)$ for $U=4.5$ (a) and $U=5.0$ (b) with various values of $\beta$.}
\end{figure}
Although the investigation on spin susceptibility and spin correlation length in the previous works with D$\Gamma$A\cite{TSchaefer2015,AAKatanin2009} and LDFA\cite{JOtsuki2014,TRibic2018} have already revealed that the low-temperature behavior of the spin fluctuations of the half-filled Hubbard model on a square lattice is consistent with the non-linear $\sigma$ model, the connection between the spin fluctuation and the pseudogap formation is still not well understood. In the previous study with the non-linear $\sigma$ model approach\cite{KBorejsza2004}, it is argued that there is a finite critical value of $U$ ($U_{\rm c}\approx 4.25$) which separates a pseudogap phase and a Mott insulating phase. In the pseudogap phase, finite $\rho(\omega=0)$ lingers at finite temperature, whereas  clear gap opening occurs in the Mott insulating phase.  The purpose of this section is to verify whether such abrupt change in the temperature dependence of $\rho(\omega=0)$ occurs at a finite $U$ or not.

To see the temperature dependence of DOS, in Fig.~\ref{DOS} those with various values of $\beta$ for $U=4.5$ and $U=5.0$ are depicted in panels (a) and (b), respectively.
Each of these DOSs consists of a peak at the Fermi level flanked by the shoulder structures corresponding to the lower- and upper-Hubbard bands at high temperatures. The pseudogap appears at a temperature close to $T_{\rm N}^{\rm DMFT}$ and further develops as temperature decreases.
While finite DOS in the vicinity of the Fermi level persists for $U=4.5$ even at the lowest temperature indicated in the figure, clear opening of the gap can be seen at $\beta=9.0$ for $U=5.0$. From these results it is expected that a pseudogap phase to Mott-Hubbard insulator transition or crossover exists in between $U=4.5$ and $U=5.0$.

\begin{figure}
\includegraphics[width=8cm]{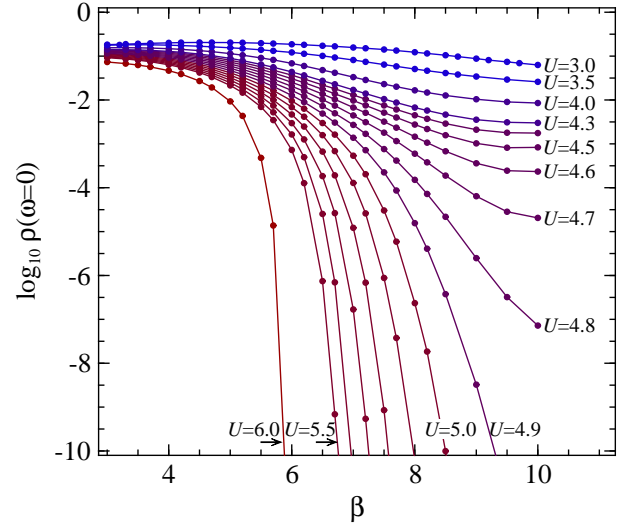}
\caption{\label{RHO}Logarithmic plots of $\rho(\omega=0)$ as functions of $\beta$ for various values of $U$: $U=3.0$, 3.5, 4.0, 6.0 and 0.1 interval from 4.3 to 5.5.}
\end{figure}
To examine how the temperature dependence of $\rho(\omega=0)$ varies as $U$ changes, in Fig.~\ref{RHO} the logarithmic plots of $\rho(\omega=0)$ as a function of $\beta$ for various values of $U$ are presented. The value of $\rho(\omega=0)$ is reduce as temperature decreases and the reduction becomes steeper as $U$ increases particularly at low temperatures; whereas the value of $\rho(\omega=0)$ for $U=4.9$ is rapidly reduced with decreasing temperature and the value is less than $10^{-10}$ at $\beta=10.0$, the reduction is moderate for $U=4.3$ and the value is only reduced to about 47\% from $\beta=3.0$ to $\beta=10.0$. 

\begin{figure}
\includegraphics[width=8cm]{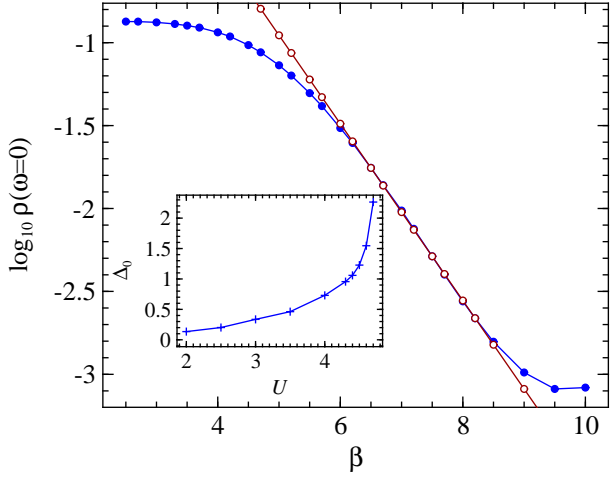}
\caption{\label{RHOU45}Logarithmic plot of  $\rho(\omega=0)$ as a function of $\beta$ for $U=4.5$ (closed circles) and its linear least squares fit within a range from $\beta=6.2$ to 8.2 (open circles). The inset shows $\Delta_0$ evaluated from the least squares analysis for various values of $U$.}
\end{figure}
In the non-linear $\sigma$ model approach\cite{KBorejsza2004}, the temperature dependence of $\rho(\omega=0)$ for the weak coupling limit at low temperatures can be approximated by 
\begin{align}
\rho(\omega=0)\propto \exp(-\beta\Delta_0), \label{rho0}
\end{align}
where $\Delta_0$ is half the size of the gap of the AFM state at $T=0$. 
To verify whether the low-temperature behavior of our results are consistent with the pseudogap phase in the non-linear $\sigma$ model approach for small $U$, the linear least squares fit is made for $\log_{10}\rho(\omega=0)$ as a function of $\beta$ to determine $\Delta_0$. For example, the results of the linear least squares fit for $U=4.5$ is shown in Fig.~\ref{RHOU45}. The fitting is made within a range from $\beta=6.2$ to 8.2. A reasonably good agreement can be obtained within the range. However, the substantial deviation from the linear approximation are found for $\beta \ge 9.0$ and the values of $\rho(\omega=0)$ are larger than those expected from Eq.~(\ref{rho0}). This is probably caused by inaccuracy due to small $1-\lambda_{\rm SP} \lesssim 5\times 10^{-4}$ for $\beta > 8.5$ as mentioned in Sec.~\ref{ssBench}. The similar tendency is found for $U\le 4.7$ at low temperatures. The estimated $\Delta_0$'s for various values of $U$ are depicted in the inset. The obtained $2\Delta_0$ is, indeed, about the peak to peak distance of the pseudogap structure of DOS for $U\le 4.3$. However, $\Delta_0$ rapidly deviates from the actual size of the pseudogap of DOS for $U> 4.3$ and the possible range of $\beta$ for the linear fitting is reduced as $U$ increases, indicating rapid disappearance of states inside the gap.  No reasonable fitting is available for $U\ge 4.8$. 

These results show that a sharp crossover from the pseudogap phase to the Mott insulator takes place at $U^{*}\approx 4.7$. Indeed, the value of $U^{*}$ coincides with the boundary between the Slater and Mott-Heisenberg regimes defined by the inflection point of $D$ curve as a function of $U$, i.e., $(\partial^2 D/\partial U^2)_{T}=0$, discussed in Sec.~\ref{ssD}. 

The robustness of the results obtained here has been checked by altering the part of the procedure of MEM described in Sec.~\ref{MEM} in several different ways, e.g., by taking data points only on the imaginary axis instead of those in Fig.\,\ref{stadium} or replacing the definition of $\chi^2$ in Eq.~(\ref{chi2}) by that with absolute errors. Although the accuracy of the data is reduced and discernible fluctuations of the data can be observed when the same $\log_{10}\rho(\omega=0)$ plots as in Fig.~\ref{RHO} are made, the variation of the estimated values of $U^{*}$ is within 0.1.

\subsection{Pseudogap in the spectral function\label{ssPG}}
In the previous study with D$\Gamma$A\cite{TSchaefer2015,GRohringer2016}, the MIT in the half-filled Hubbard model on a square lattice has been discussed, where the transition temperature is related to the variation in the $\omega_n$ dependence of ${\rm Im}\,\Sigma_{\bm{k}}(i\omega_n)$. If the system is a Fermi liquid or at least a Fermi liquid like, it is expected that $|{\rm Im}\,\Sigma_{\bm{k}}(\omega)|$ should decrease as $\omega$ decreases. However, the loss of the Fermi-liquid feature is not necessarily indicate insulating behavior of the system. The purpose of this subsection is to clarify the relation between the pseudogap formation in the spectral function $A_{\bm{k}}(\omega)$ and the change in ${\rm Im}\,\Sigma_{\bm{k}}(i\omega_n)$.

\begin{figure}
\includegraphics[width=8cm]{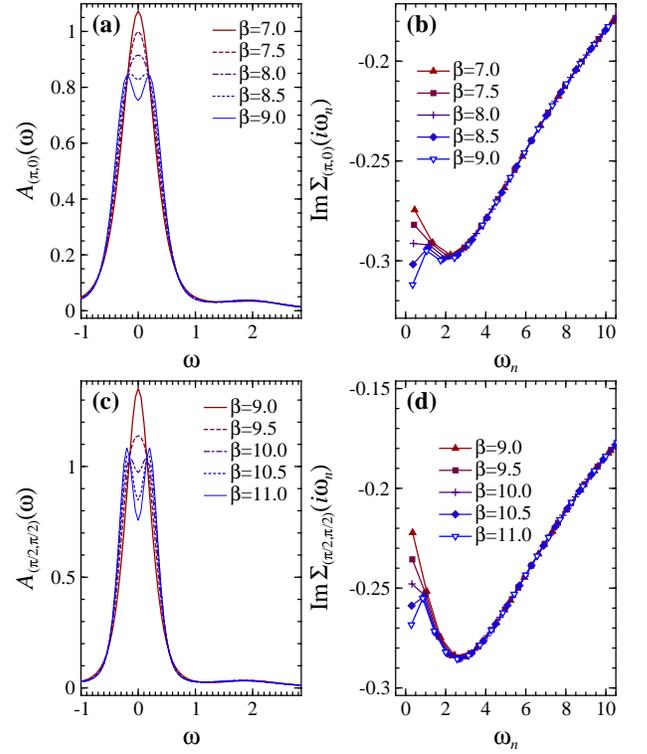}
\caption{\label{SGMkU30}Relationship between the pseudogap formation at the Fermi level of $A_{\bm{k}}(\omega)$ for $U=3.0$ and corresponding ${\rm Im}\,\Sigma_{\bm{k}}(i\omega_n)$. $A_{\bm{k}}(\omega)$ at the X point [$\bm{k}=(\pi,0)$] in the vicinity of the Fermi level with temperatures across the pseudogap formation $\beta=7.0$, 7.5, 8.0, 8.5, and 9.0 are shown in (a) and corresponding ${\rm Im}\,\Sigma_{\bm{k}}(i\omega_n)$ on the imaginary axis are depicted in (b). Those in (c) and (d) are the same as in (a) and (b) but for the M$_2$ point [$\bm{k}=(\pi/2,\pi/2)$] with $\beta=9.0$, 9.5, 10.0, 10.5, and 11.0.}
\end{figure}
As discussed in Sec.~\ref{ssAk}, the pseudogap formation in $A_{\bm{k}}(\omega)$ has $\bm{k}$ dependence; it initiates at the X point and spread through the Fermi surface and terminates at the M$_2$ point as temperature decreases. Figure~\ref{SGMkU30} shows $A_{\bm{k}}(\omega)$ in the vicinity of the Fermi level for $U=3.0$ at temperatures across the pseudogap formation at the X and M$_2$ points together with the corresponding ${\rm Im}\,\Sigma_{\bm{k}}(i\omega_n)$. It is clearly seen in panels (b) and (d) that the $\omega_n$ dependence of ${\rm Im}\,\Sigma_{\bm{k}}(i\omega_n)$ changes upturn to downturn in ${\rm Im}\,\Sigma_{\bm{k}}(i\omega_n)$ with decreasing temperature around the pseudogap formation temperature. These results are in good agreement with those of D$\Gamma$A calculations for $U=3.0$ in Ref.~\onlinecite{GRohringer2016}.
This indicates that for small $U$ the Fermi-liquid feature is gradually lost accompanied by the pseudogap formation in $A_{\bm{k}}(\omega)$ at the Fermi level. It is also found in panels (b) and (d) that the variation mainly occurs in  the lowest Matsubara frequency $\omega_1=\pi/\beta$, where $|{\rm Im}\,\Sigma_{\bm{k}}(i\pi/\beta)|$ is increases with decreasing temperature. This predominant increase in $|{\rm Im}\,\Sigma_{\bm{k}}(i\pi/\beta)|$, as was already pointed out in Ref.~\onlinecite{YMVilk1996}, is the main cause of the double peak structure in $A_{\bm{k}}(\omega)$, i.e., the formation of the pseudogap. Since the low-temperature magnetic excitation of the half-filled Hubbard model on a square lattice is considered to be described by the two-dimensional non-linear $\sigma$ model in the renormalized {\it classical} regime\cite{SChakravarty1988}, the magnetic scattering at the lowest Matsubara frequency is expected to be the dominant process, which is consistent with our results. 

\begin{figure}
\includegraphics[width=8cm]{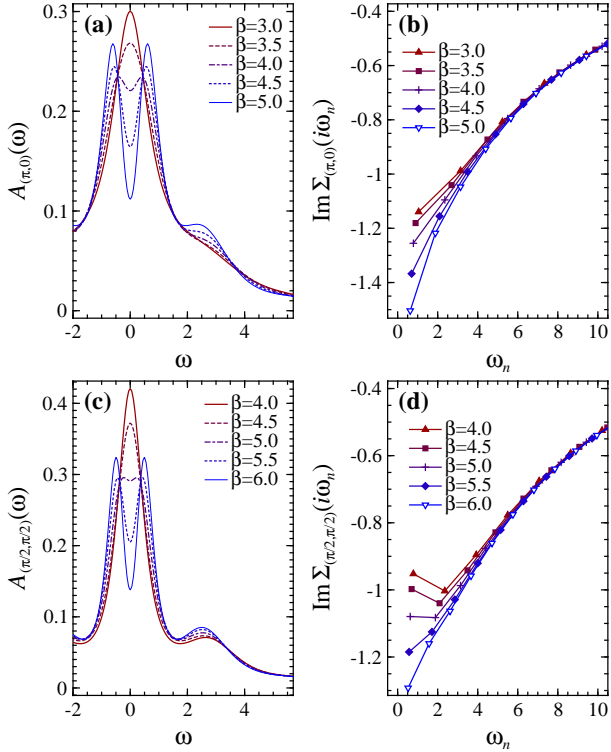}
\caption{\label{SGMkU50}The same as Fig.\,\ref{SGMkU30} but for $U=5.0$.}
\end{figure}
The relation, however, becomes less clear as $U$ increases, although the pseudogap formation still occurs below $T_{\rm N}^{\rm DMFT}$.  As shown in Fig.~\ref{SGMkU50}, for $U=5.0$, $|{\rm Im}\,\Sigma_{\bm{k}}(i\omega_n)|$ at the X point monotonically increases with decreasing $\omega_n$ already at high temperatures above $T_{\rm N}^{\rm DMFT}$ and therefore no upturn to downturn change occurs in ${\rm Im}\,\Sigma_{\bm{k}}(i\omega_n)$ at $\beta=4.75$ where the pseudogap appears.  This non-Fermi-liquid feature in ${\rm Im}\,\Sigma_{\bm{k}}(i\omega_n)$ above $T_{\rm N}^{\rm DMFT}$ spreads through the Fermi surface from the X to the M$_2$ point as $U$ increases from $U=4.0$ and at $U=5.5$ the whole Fermi surface is lost already above the temperature where the pseudogap appears, resulting in no upturn to downturn change in ${\rm Im}\,\Sigma_{\bm{k}}(i\omega_n)$ at all.  

Furthermore, one can see the enhancement of the intensity of shoulder structures around $\omega\approx\pm 2.5$ in $A_{\bm{k}}(\omega)$ and the increase of $|{\rm Im}\,\Sigma_{\bm{k}}(i\omega_n)|$ up to $\omega_n \sim 5$ with decreasing temperature. 
The appearance of the high-energy structure $\omega\approx\pm U/2$ is hallmark of the Mott physics and cannot be explained by the magnetic scattering within the energy scale of $T_{\rm N}^{\rm DMFT}\sim 0.3$. These facts can be contrasted with $A_{\bm{k}}(\omega)$ and ${\rm Im}\,\Sigma_{\bm{k}}(i\omega_n)$ for $U=3.0$ in Fig.~\ref{SGMkU30}, where their temperature effects are mainly found within the energy scale of $T_{\rm N}^{\rm DMFT}\sim 0.2$ around the Fermi level and are in accordance with the Slater mechanism of gap formation due to the short-range AFM ordering. Nevertheless, a quasiparticle-like broad single peak still exists at the Fermi level at high temperatures in $A_{\rm k}(\omega)$ for $U=5.0$  in Fig.~\ref{SGMkU50} and predominant increase in $|{\rm Im}\,\Sigma_{\bm{k}}(i\pi/\beta)|$ owing to the magnetic scattering leads to the pseudogap formation.

\subsection{$U$--$T$ phase diagram\label{ssUT}}
To conclude this section, here we consider how the electronic state changes on the $U$-$T$ parameter space. In Fig.~\ref{UTphase}, characteristic temperatures so far discussed are summarized. The $D$ local maximum temperature is the onset temperature of AFM correlation in the high-temperature metallic state, which is signaled by increase in $|{\rm Im}\,\Sigma_{\bm{k}}(i\pi/\beta)|$ with decreasing temperature around the X point for $U< 5.0$. For $U\ge 5.5$, on the other hand, strong correlations already develop at high-temperatures and no $D$ local maximum temperature found below $T < 0.5$. The AFM correlations further develop with decreasing temperature and at $T^{\rm DMFT}_{\rm N}$, $|{\rm Im}\,\Sigma_{\bm{k}}(i\pi/\beta)|$ with momenta $\bm{k}$ on whole Fermi surface start to increase with decreasing temperature.

For $U \le 3.5$, the Fermi-liquid behavior is gradually lost below $T^{\rm DMFT}_{\rm N}$ line as the AFM correlations develops with decreasing temperature. This starts at the X point of the Fermi surface and ends at the M$_2$ point with concomitant formation of the pseudogap in the corresponding quasiparticle peak at the Fermi level. The Fermi surface is totally lost and the formation of the pseudogap at the Fermi level is complete at the M$_2$ pseudogap line. The difference between the pseudogap temperatures at the X and M$_2$ points  rapidly decreases as $U$ is reduced and is only about 2$\times10^{-3}$ at $U=2.0$.

The same gradual pseudogap formation in the peak at the Fermi level occurs for $U \ge 4.0$ below the $T^{\rm DMFT}_{\rm N}$ line. In this case, however, $|{\rm Im}\,\Sigma_{\bm{k}}(i\pi/\beta)|$ around the X point is large even at high temperatures and the system is already non-Fermi-liquid like above the pseudogap formation temperature. This tendency is strengthen as $U$ increases and for $U \ge 5.5$, the whole Fermi surface is already non-Fermi-liquid like above the pseudogap formation temperature, where $A_{\bm{k}}(\omega)$ has a structure with a broad single peak at the Fermi level accompanied by prominent shoulder structures at $\omega\approx\pm U/2$.  In this regards, the formation of the pseudogap at low temperatures only heralds the development of AFM correlations for $U \ge 5.5$ and local spins are considered to be preformed at higher temperatures as expected in the Mott-Heisenberg regime. This is contrasted with the simultaneous formation of the pseudogap and loss of the Fermi-liquid feature found for $U \le 3.5$, which is expected in the Slater regime.  Note that the $D$ local maximum and the X and M$_2$ pseudogap lines are only those of crossover and no anomaly which indicates a true transition is found in $D$, DOS or the spectral functions.

Further lowering the temperature, the development of the pseudogap is distinctly different depending on whether $U$ is larger or smaller than $U^{*} \approx 4.7$. For $U < U^{*}$, $\rho(\omega=0)$ is reduced with decreasing temperature but persists even at low temperatures. In contrast, for $U > U^{*}$, the reduction is much rapid and clear gap opening occurs at certain temperature. The value of $U^{*} \approx 4.7$ coincides with the boundary between the Slater and Mott-Heisenberg regimes defined by the inflection point of $D$ curve as a function of $U$.

\section{Discussions\label{Discussions}}
The MIT in the half-filled Hubbard model on the square lattice has been discussed using D$\Gamma$A and lattice QMC\cite{TSchaefer2015}. Although the difference between the pseudogap formation temperatures at the X and M$_2$ points is substantially larger compared to the present study,  the behavior of the formation found in this D$\Gamma$A study for $U \le 4.0$ is essentially the same to the present study (see Fig.\,\ref{UTphase}): the pseudogap formation starts at the X point and ends at the M$_2$ point with decreasing temperature. From this results it is suggested that $U_{\rm c}=0$ for $T\to 0$ and thus no MIT occurs at any $U>0$ similar to the one-dimensional Hubbard model in this D$\Gamma$A study. Although we found a sharp crossover from the pseudogap phase to Mott insulator around $U^{*}\approx 4.7$, we did not find any anomaly, e.g., a discontinuity or kink, in $D$ up to the second derivative with respect to $U$ at $U^{*}$. Hence, it is unlikely that $U^{*}\approx 4.7$ is the true MIT point. For this reason, in the strict sense, the true MIT point is considered to be presence at $U_{\rm c}=0$.  This weakness of the variation from the metallic to insulating phase at $U^{*}$ as compared to that in infinite dimension is not surprising.  Since the pseudogap owing to the AFM correlation already exists at higher temperatures, the crossover we have discussed here amounts to a subtle change in the states in the gap. This is contrasted with the MIT in the Hubbard model in infinite dimension, where the abrupt destruction of the coherent peak at the Fermi level causes the first-order MIT in finite temperature.  

In the D$\Gamma$A study, there is no detailed investigation on DOS at the Fermi level  below the pseudogap formation temperature in particular for $U>4.0$. For this reason, there is a possibility that a similar sharp crossover from the Slater to Mott-Heisenberg regime is found at finite $U$ in D$\Gamma$A and it would be interesting to investigate DOS at the Fermi level below the pseudogap formation temperature with  D$\Gamma$A to clarify this point.

\section{Conclusions\label{Conclusions}}
A new formula for the two-body Green's function combined with the Lanczos ED technique proposed in this paper provides efficient and accurate means to calculate the local vertex function of the effective impurity Anderson model required for DFA and similar perturbative extensions of DMFT. The formula is applicable to not only the impurity Anderson model but cluster models including those with multiple orbitals. Utilizing this new scheme, the double occupancy, kinetic energy, spectral function and DOS of the Hubbard model on the square lattice at the half-filling are calculated by means of LDFA with an unprecedented accuracy at low temperatures and $U$'s ranging from $U=2.0$ to 6.0 to discuss the metal-insulator transition. 

It is found that the pseudogap is first formed in the quasiparticle peak of the spectral function at the X point as temperature decreases. The formation spreads through the Fermi surface and ends at the M$_2$ point $\bm{k}=(\pi/2,\pi/2)$ similar to the previous D$\Gamma$A and lattice QMC study\cite{TSchaefer2015}.  For $U \le 3.5$, the pseudogap formation and the loss of the Fermi-liquid feature occur simultaneously  and below the temperature of the pseudogap formation at the M$_2$ point, the Fermi surface is totally lost and the system enters the pseudogap phase. These results for $U \le 3.5$ are consistent with those expected in the Slater regime. 

However, for larger $U$, although the formation of the pseudogap still occurs in the quasiparticle-like single peak at the Fermi level accompanied by prominent shoulder structures at $\omega\approx\pm U/2$, the Fermi-liquid feature is partially lost around the X point already at higher temperatures above $T^{\rm DMFT}_{\rm N}$ for $U=4.0$ and totally lost for $U\ge 5.5$. These results for $U \ge 4.0$ is consistent with those expected in the Mott-Heisenberg regime, in which local spins are preformed above the temperature where AFM correlations start to develop.

A sharp crossover from the pseudogap phase to the Mott insulator around $U^{*} \approx 4.7$ is found to occur below the temperature of the pseudogap formation. For $U < U^{*}$, $\rho(\omega=0)$ is reduced with decreasing temperature but persists even at low temperatures. In contrast, for $U > U^{*}$, the reduction is much rapid and clear gap opening occurs at certain temperature. These low-energy behavior of DOS in the vicinity of the Fermi level is consistent with the previous study with the non-linear $\sigma$ model approach\cite{KBorejsza2004}.
\begin{acknowledgements}
This work was supported by JSPS KAKENHI Grant No.~25400377, No.~18K03541, and No.~18H03683. The author also would like to thank A. Lichtenstein for fruitful discussions.
\end{acknowledgements}
\appendix
\section{Derivation of the new expression for two-body Green's function}
The purpose of this section is to provide detailed description of a derivation of the expression for the two-body Green's function given in Eqs.~(\ref{g2omega})-(\ref{g2pp}).
\begin{widetext}
\begin{align}
\chi_{1234}(i\omega_1,i\omega_2;i\omega_3,i\omega_1+i\omega_2-i\omega_3) &=
\int_0^\beta d\tau_1\int_0^\beta d\tau_2\int_0^\beta d\tau_3\langle {\cal T}[c_1(\tau_1)c_2(\tau_2)c_3^\dagger(\tau_3)c_4^\dagger(0)]\rangle e^{i(\omega_1\tau_1+\omega_2\tau_2-\omega_3\tau_3)} \nonumber\\
&=\frac{1}{Z}\sum_{lmnk}\langle k|c_1|l\rangle\langle l|c_2|m\rangle\langle m|c^\dagger_3|n\rangle\langle n|c^\dagger_4|k\rangle 
 \phi_{lmnk}(\omega_1,\omega_2,-\omega_3) \nonumber\\
&+\frac{1}{Z}\sum_{lmnk}\langle k|c^\dagger_3|l\rangle\langle l|c_1|m\rangle\langle m|c_2|n\rangle\langle n|c^\dagger_4|k\rangle 
 \phi_{lmnk}(-\omega_3,\omega_1,\omega_2) \nonumber\\
&+\frac{1}{Z}\sum_{lmnk}\langle k|c_2|l\rangle\langle l|c^\dagger_3|m\rangle\langle m|c_1|n\rangle\langle n|c^\dagger_4|k\rangle 
 \phi_{lmnk}(\omega_2,-\omega_3,\omega_1) \nonumber\\
&-\frac{1}{Z}\sum_{lmnk}\langle k|c^\dagger_3|l\rangle\langle l|c_2|m\rangle\langle m|c_1|n\rangle\langle n|c^\dagger_4|k\rangle 
 \phi_{lmnk}(-\omega_3,\omega_2,\omega_1) \nonumber\\
&-\frac{1}{Z}\sum_{lmnk}\langle k|c_1|l\rangle\langle l|c^\dagger_3|m\rangle\langle m|c_2|n\rangle\langle n|c^\dagger_4|k\rangle 
 \phi_{lmnk}(\omega_1,-\omega_3,\omega_2) \nonumber\\
&-\frac{1}{Z}\sum_{lmnk}\langle k|c_2|l\rangle\langle l|c_1|m\rangle\langle m|c^\dagger_3|n\rangle\langle n|c^\dagger_4|k\rangle 
 \phi_{lmnk}(\omega_2,\omega_1,-\omega_3) \label{g2a}
\end{align}
where 
\begin{align}
 \phi_{lmnk}(\omega_1,\omega_2,\omega_3)=e^{-\beta E_k}\int_0^\beta d\tau_1\int_0^{\tau_1} d\tau_2\int_0^{\tau_2} d\tau_3 e^{(E_k-E_l+i\omega_1)\tau_1} e^{(E_l-E_m+i\omega_2)\tau_2}e^{(E_m-E_n+i\omega_3)\tau_3}\label{intphi}
\end{align}

If both the condition $E_k \ne E_m$ or $\omega_1+\omega_2\ne 0$ and the condition  $E_l \ne E_n$ or $\omega_2 +\omega_3\ne 0$ are satisfied, we obtain
 \begin{align}
\phi_{lmnk}(\omega_1,\omega_2,\omega_3)=&-\frac{e^{-\beta E_n}-e^{-\beta E_k}}{(E_m-E_n+i\omega_3)(E_l-E_n+i(\omega_2+\omega_3))(E_k-E_n+i(\omega_1+\omega_2+\omega_3))}\nonumber\\
&+\frac{e^{-\beta E_l}-e^{-\beta E_k}}{(E_m-E_n+i\omega_3)(E_l-E_n+i(\omega_2+\omega_3))(E_k-E_l+i\omega_1)}\nonumber\\
&-\frac{e^{-\beta E_m}-e^{-\beta E_k}}{(E_m-E_n+i\omega_3)(E_l-E_m+i\omega_2)(E_k-E_m+i(\omega_1+\omega_2))}\nonumber\\
&-\frac{e^{-\beta E_l}-e^{-\beta E_k}}{(E_m-E_n+i\omega_3)(E_l-E_m+i\omega_2)(E_k-E_l+i\omega_1)}.\label{phi}
\end{align}
Now we rewrite the right-hand side of Eq.(\ref{phi}) in such a way that the three factors in denominator and the argument of the exponential function of each term contain the same eigenvalue of ${\cal H}$. To do this, we use the identity
\begin{align}
\frac{1}{z_1z_2}=\frac{1}{z_2\pm z_1}\left(\frac{1}{z_1}\pm \frac{1}{z_2}\right).\label{z1z2}
\end{align}
For instance, the term with the factor $e^{-\beta E_l}$ on the second line of Eq.~(\ref{phi}) can be transformed into
\begin{align}
e^{-\beta E_l}\left(\frac{1}{E_m-E_n+i\omega_3}-\frac{1}{E_l-E_n+i(\omega_2+\omega_3)}\right)\frac{1}{(E_l-E_m+i\omega_2)(E_k-E_l+i\omega_1)}.
\end{align}
Together with the term with the factor $e^{-\beta E_l}$ on the fourth line of Eq.(\ref{phi}), we get
\begin{align}
-e^{-\beta E_l}\frac{1}{(E_l-E_n+i(\omega_2+\omega_3))(E_l-E_m+i\omega_2)(E_k-E_l+i\omega_1)},
\end{align}
where all the factors in the denominator and $e^{-\beta E_l}$ contain the same eigenvalue $E_l$.
We can cast all the terms of Eq.~(\ref{phi}) in this form by further repeated use of Eq.~(\ref{z1z2}) to the terms with $e^{-\beta E_k}$:
\begin{align}
\phi_{lmnk}(\omega_1,\omega_2,\omega_3)=&-e^{-\beta E_k}\frac{1}{(E_k-E_l+i\omega_1)(E_k-E_m+i(\omega_1+\omega_2))(E_k-E_n+i(\omega_1+\omega_2+\omega_3))}\nonumber\\
&-e^{-\beta E_m}\frac{1}{(E_m-E_n+i\omega_3)(E_m-E_k-i(\omega_1+\omega_2))(E_m-E_l-i\omega_2)}\nonumber\\
&+e^{-\beta E_l}\frac{1}{(E_l-E_m+i\omega_2)(E_l-E_n+i(\omega_2+\omega_3))(E_l-E_k-i\omega_1)}\nonumber\\
&+e^{-\beta E_n}\frac{1}{(E_n-E_k-i(\omega_1+\omega_2+\omega_3))(E_n-E_l-i(\omega_2+\omega_3))(E_n-E_m-i\omega_3)}.\label{phi2}
\end{align}
These terms correspond to the major terms discussed in Sec.~\ref{Expression} such as Eq.~(\ref{major}).

We now deal with the situations with $E_k=E_m$ or $E_l=E_n$, where some of denominators in Eq.~(\ref{phi2}) are zero if $\omega_1+\omega_2= 0$ or $\omega_2+\omega_3= 0$ is satisfied.
When $E_k=E_m$ and $\omega_1+\omega_2\ne 0$, using Eq.~(\ref{z1z2}) the first two lines on the right-hand side of Eq.(\ref{phi2}) can be written as
\begin{align}
&-e^{-\beta E_k}\left\{\frac{1}{(E_k-E_l+i\omega_1)i(\omega_1+\omega_2)(E_k-E_n+i(\omega_1+\omega_2+\omega_3))}
-\frac{1}{(E_k-E_n+i\omega_3)i(\omega_1+\omega_2)(E_k-E_l-i\omega_2)}\right\}\nonumber  \\
&=-e^{-\beta E_k}\Big[\frac{1}{E_k-E_l+i\omega_1}\frac{1}{E_k-E_n+i\omega_3}\left\{ \frac{1}{i(\omega_1+\omega_2)}-\frac{1}{E_k-E_n+i(\omega_1+\omega_2+\omega_3)}\right\}\nonumber  \\
&~~~~~~~~~~~~-\frac{1}{E_k-E_n+i\omega_3}\frac{1}{E_k-E_l+i\omega_1}\left\{\frac{1}{i(\omega_1+\omega_2)}+\frac{1}{E_k-E_l-i\omega_2}\right\}  \Big]  \nonumber  \\
&=e^{-\beta E_k}\frac{1}{(E_k-E_l+i\omega_1)(E_k-E_n+i\omega_3)}\left\{\frac{1}{E_k-E_n+i(\omega_1+\omega_2+\omega_3)}
+\frac{1}{E_k-E_l-i\omega_2}\right\}\label{phi2b}
\end{align}
Similarly, when $E_l=E_n$ and $\omega_2+\omega_3\ne 0$, the last two lines on the right-hand side of Eq.(\ref{phi2}) can be written as
\begin{align}
&e^{-\beta E_l}\left\{\frac{1}{(E_l-E_m+i\omega_2)i(\omega_2+\omega_3)(E_l-E_k-i\omega_1)}
                                     -\frac{1}{(E_l-E_k-i(\omega_1+\omega_2+\omega_3))i(\omega_2+\omega_3)(E_l-E_m-i\omega_3)}\right\} \nonumber  \\
&=-e^{-\beta E_l}\frac{1}{(E_l-E_k-i(\omega_1+\omega_2+\omega_3))(E_l-E_m+i\omega_2)}\left\{\frac{1}{ E_l-E_k-i\omega_1}+\frac{1}{E_l-E_m-i\omega_3}\right\}\label{phi2c}
\end{align}
To complete the calculation of $\phi$, we further need to know special cases of the integration of Eq.~(\ref{intphi}).
If $E_k=E_m$ and $\omega_1+\omega_2=0$, we obtain
\begin{align}
\phi_{lmnk}(\omega_1,\omega_2,\omega_3)=&+e^{-\beta E_k}\frac{1}{(E_k-E_l+i\omega_1)(E_k-E_n+i\omega_3)}\left\{\frac{1}{E_k-E_n+i\omega_3}+\frac{1}{E_k-E_l+i\omega_1}\right\} \nonumber \\
&+e^{-\beta E_l}\frac{1}{(E_l-E_k-i\omega_1)^2(E_l-E_n+i(\omega_2+\omega_3))}\nonumber\\
&+e^{-\beta E_n}\frac{1}{(E_n-E_k-i\omega_3)^2(E_n-E_l-i(\omega_2+\omega_3))}\nonumber\\
&+\beta e^{-\beta E_k}\frac{1}{(E_k-E_n+i\omega_3)(E_k-E_l-i\omega_2)}.\label{phi3a}
\end{align}
The terms on the first line of Eq.~(\ref{phi3a}) are identical to Eq.~(\ref{phi2b}) with $E_k=E_m$ and $\omega_1+\omega_2=0$ and the same holds for the terms on the second and third lines of Eq.~(\ref{phi3a}), which correspond to the terms on the last two lines of  Eq.~(\ref{phi2}). However the term on the last line of Eq.~(\ref{phi3a}) is newly appeared.
Similarly, if $E_l=E_n$ and $\omega_2+\omega_3=0$, we obtain
\begin{align}
\phi_{lmnk}(\omega_1,\omega_2,\omega_3)=&-e^{-\beta E_k}\frac{1}{(E_k-E_l+i\omega_1)^2(E_k-E_m+i(\omega_1+\omega_2))}\nonumber \\
&-e^{-\beta E_m}\frac{1}{(E_m-E_l-i\omega_2)^2(E_m-E_k-i(\omega_1+\omega_2))}\nonumber \\
&-e^{-\beta E_l}\frac{1}{(E_l-E_k-i\omega_1)(E_l-E_m+i\omega_2)}\left\{\frac{1}{E_l-E_k-i\omega_1}+\frac{1}{E_l-E_m+i\omega_2}\right\}\nonumber\\
&-\beta e^{-\beta E_l}\frac{1}{(E_l-E_k-i\omega_1)(E_l-E_m+i\omega_2)}.\label{phi3b}
\end{align}
We find  the terms on the third line of Eq.~(\ref{phi3b}) are identical to Eq.~(\ref{phi2c}) with $E_l=E_n$ and $\omega_2+\omega_3=0$ and the terms on the first two lines of Eq.~(\ref{phi3b}) correspond to those on the first two lines of  Eq.~(\ref{phi2}). Again, we see the additional term on the last line of Eq.~(\ref{phi3b}). From Eqs.~(\ref{phi2})-(\ref{phi3b}), we obtain
\begin{align}
&\phi_{lmnk}(\omega_1,\omega_2,\omega_3)=\nonumber\\
&-(1-\delta_{E_k,E_m})e^{-\beta E_k}\frac{1}{(E_k-E_l+i\omega_1)(E_k-E_m+i(\omega_1+\omega_2))(E_k-E_n+i(\omega_1+\omega_2+\omega_3))}\nonumber\\
&-(1-\delta_{E_k,E_m})e^{-\beta E_m}\frac{1}{(E_m-E_n-i\omega_3)(E_m-E_k-i(\omega_1+\omega_2))(E_m-E_l-i\omega_2)}\nonumber\\
&+(1-\delta_{E_l,E_n})e^{-\beta E_l}\frac{1}{(E_l-E_m+i\omega_2)(E_l-E_n+i(\omega_2+\omega_3))(E_l-E_k-i\omega_1)}\nonumber\\
&+(1-\delta_{E_l,E_n})e^{-\beta E_n}\frac{1}{(E_n-E_k-i(\omega_1+\omega_2+\omega_3))(E_n-E_l-i(\omega_2+\omega_3))(E_n-E_m-i\omega_3)}\nonumber\\
&+\delta_{E_k,E_m}e^{-\beta E_k}\frac{1}{(E_k-E_l+i\omega_1)(E_k-E_n+i\omega_3)}\left\{\frac{1}{E_k-E_n+i(\omega_1+\omega_2+\omega_3)}
+\frac{1}{E_k-E_l-i\omega_2}\right\}\nonumber\\
&-\delta_{E_l,E_n}e^{-\beta E_l}\frac{1}{(E_l-E_k-i(\omega_1+\omega_2+\omega_3))(E_l-E_m+i\omega_2)}\left\{\frac{1}{ E_l-E_k-i\omega_1}+\frac{1}{E_l-E_m-i\omega_3}\right\}\nonumber\\
&+\delta_{E_k,E_m}\delta_{\omega_1,-\omega_2}\beta e^{-\beta E_k}\frac{1}{(E_k-E_n+i\omega_3)(E_k-E_l-i\omega_2)}
-\delta_{E_l,E_n}\delta_{\omega_2,-\omega_3}\beta e^{-\beta E_l}\frac{1}{(E_l-E_k-i\omega_1)(E_l-E_m+i\omega_2)}\label{phi4}
\end{align}
\end{widetext}
Finally, inserting Eq.~(\ref{phi4}) into Eq.~(\ref{g2a}), we obtain the expression for the two-body Green's function  in Eqs.~(\ref{g2omega})-(\ref{g2pp}).

\section{Derivation of update formula of hybridization function}
Here, we show the derivation of the update formula of $\Delta_\omega$ for DFA in Eq.~(\ref{deltanewdfa}).
For the DMFT calculation, one can choose new $\Delta_\omega$ and  $g_\omega$ for next iteration
\begin{align}
g_{\omega}^{\rm new}&=\langle g_{\bm{k}\omega} \rangle_{\bm{k}}, \label{gnew}\\
\Delta_{\omega}^{\rm new}&=g_{\omega}^{-1}-[g^{\rm new}_{\omega}]^{-1}+\Delta_\omega,\label{deltanew}
\end{align}
where $g_{\bm{k}\omega}\equiv [g_{\omega}^{-1}+\Delta_\omega-\varepsilon_{\bm{k}}]^{-1}$ is the DMFT lattice Green's function. 
These update formulas render robust and rapid convergence of $\Delta_\omega$ for DMFT. A formula similar to Eq.~(\ref{gnew}) can be derived for DFA. The condition of the convergence of $\Delta_\omega$ adopted in this study is $\langle G^d_{\bm{k}\omega}\rangle_{\bm{k}}=0$.
We rewrite this equation as
\begin{align}
&\langle G^d_{\bm{k}\omega}\rangle_{\bm{k}}\nonumber \\
&=\left\langle G^d_{\bm{k}\omega} [G^{d,0}_{\bm{k}\omega}]^{-1}G^{d,0}_{\bm{k}\omega}\right\rangle_{\bm{k}} \nonumber \\
&=\left\langle G^d_{\bm{k}\omega} [G^{d,0}_{\bm{k}\omega}]^{-1}(-g_\omega+g_{\bm{k}\omega})\right\rangle_{\bm{k}} \nonumber \\
&=-\left\langle G^d_{\bm{k}\omega} [G^{d,0}_{\bm{k}\omega}]^{-1}\right\rangle_{\bm{k}}g_\omega+\left\langle G^d_{\bm{k}\omega} [G^{d,0}_{\bm{k}\omega}]^{-1}g_{\bm{k}\omega}\right\rangle_{\bm{k}}=0
\end{align}
From this one may employ
\begin{align}
g^{\rm new}_\omega=\left\langle G^d_{\bm{k}\omega} [G^{d,0}_{\bm{k}\omega}]^{-1}\right\rangle_{\bm{k}}^{-1}\left\langle G^d_{\bm{k}\omega} [G^{d,0}_{\bm{k}\omega}]^{-1}g_{\bm{k}\omega}\right\rangle_{\bm{k}} \label{gnewdfa}
\end{align}
as an update for $g_\omega$ of DFA. Indeed, if $\Sigma^d_\omega=0$, this equation is reduced to Eq.~(\ref{gnew}). With combined use of Eq.~(\ref{deltanew}), one can obtain new hybridization function for DFA. Finally, substitution of Eq.~(\ref{gnewdfa}) into Eq.~(\ref{deltanew}) and use of the relation $g_{\bm{k}\omega}^{-1}=g_{\omega}^{-1}+\Delta_\omega-\varepsilon_{\bm{k}}$ results in  Eq.~(\ref{deltanewdfa}). Similar formula can be obtained for DMFT: $\Delta_\omega^{\rm new}=g_\omega^{-1}\langle g_{\bm{k}\omega}\varepsilon_{\bm{k}}\rangle_{\bm{k}}$. Note that instead of directly calculating $G^d_{\bm{k}\omega}[G^{d,0}_{\bm{k}\omega}]^{-1}$ in Eq.~(\ref{deltanewdfa}), it is preferable to use
\begin{align}
G^d_{\bm{k}\omega}[G^{d,0}_{\bm{k}\omega}]^{-1}=[1-\Sigma^d_{\bm{k}\omega}G^{d,0}_{\bm{k}\omega}]^{-1}
\end{align}
to avoid loss of significant digits with small interaction.

\section{Convergence of the vertex function of IAM with the Lanczos ED method}
\begin{figure}
\includegraphics[width=8cm]{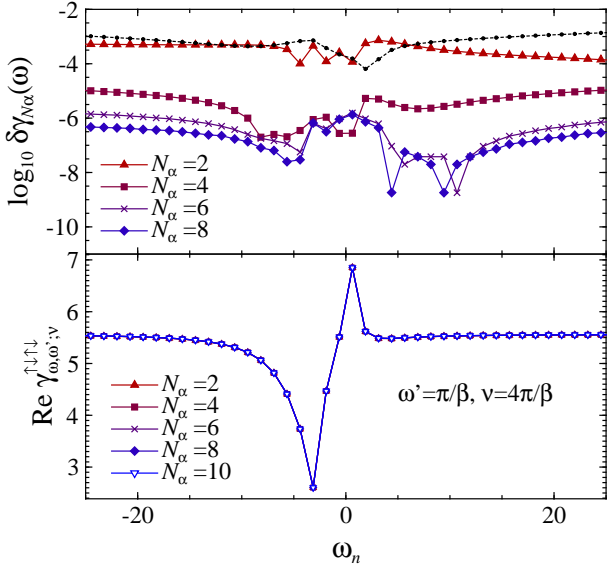}
\caption{\label{vtx4conv}Comparison of the real part of the local four-point vertex function of the IAM obtained with various values of $N_\alpha$. In lower panel, Re\,$\gamma^{\uparrow\downarrow\uparrow\downarrow}_{\omega\omega';\nu}$ as a function of the Matsubara frequency $\omega_n$ is shown for $\omega'=\pi/\beta$ and $\nu=4\pi/\beta$. In upper panel, logarithmic plots of the relative difference between Re\,$\gamma^{\uparrow\downarrow\uparrow\downarrow}_{\omega\omega';\nu}$ obtained with $N_\alpha$\,(=2, 4, 6 and 8) and that of $N_\alpha=10$ are presented [see Eq.~(\ref{dgamma})]; The dashed line is the same as $N_\alpha\,=6$ but $\Omega_6$ is omitted in the calculation.} 
\end{figure}
As described in Sec.~\ref{LanczosED}, for the left and right resolvents of the major terms, it is clear that both the low energy properties and asymptotic behavior can be accurately captured within the Lanczos scheme. For the resolvent in the center, however, excitations through the left and right resolvents must be considered and in the present method those at the finite number of the reference energy points $\Omega_{\alpha}$ ($\alpha=1,2,\cdots, N_\alpha$) are taken into account as the initial vectors of the band Lanczos method to generate the basis set to construct approximated resolvent [see Eq.\,(\ref{exvectors})]. It is, therefore, important to know how many energy points are required to obtain well converged results within the present  Lanczos algorithm. In this section, the convergence of the local four-point vertex function of the IAM with respect to the number of the energy points $N_\alpha$ is discussed as an example.

In the calculation, the IAM with $N_{\rm b}=7$ discretized conduction band levels in Eq.\,(\ref{IAM}) is assumed.
For the parameters of the IAM, $U=4$, $\beta=5$ and $\mu=2$ are adopted and the energies $\varepsilon^{\rm b}_l$ and hybridization strength $V_l$ of the discretized levels are $\varepsilon^{\rm b}_{1,7}=\pm 6$ with $V_{1,7}=0.55$, $\varepsilon^{\rm b}_{2,6}=\pm 3$ with $V_{2,6}=0.9$, $\varepsilon^{\rm b}_{3,5}=\pm 1$ with $V_{3,5}=0.85$ and  $\varepsilon^{\rm b}_4=0$ with $V_4=0.6$. These are about the values of the effective IAM of the DFA calculation of the 2D Hubbard model at half-filling with $U=4$, $\beta=5$. The reference energy points chosen are 
$\Omega_1=0$, $\Omega_{N\alpha}=5.12W$ and 
\begin{align}
\Omega_\alpha=0.02\times 2^{(\alpha-2)}W~~(\alpha=2,3,\cdots,N_\alpha-1),
\end{align}
where $W=8.94$. $\gamma_{\alpha}=0.1W$ for all $\alpha$.

In lower panel of Fig.~\ref{vtx4conv}, the real part of $\gamma^{\uparrow\downarrow\uparrow\downarrow}_{\omega\omega';\nu}$ [refer Eqs.\,(\ref{vtx4def}) and (\ref{vtx4note}) for the definition] as a function of the Matsubara frequency $\omega_n$ is depicted for $\omega'=\pi/\beta$ and $\nu=4\pi/\beta$.
The convergence is very rapid and already Re\,$\gamma^{\uparrow\downarrow\uparrow\downarrow}_{\omega\omega';\nu}$ obtained with $N_\alpha=2$ is hardly distinguishable from the others with larger $N_\alpha$. To examine the accuracy of the results more closely, the relative differences between Re\,$\gamma^{\uparrow\downarrow\uparrow\downarrow}_{\omega\omega';\nu}$ obtained with $N_\alpha$\,(=2, 4, 6 and 8) and that of $N_\alpha=10$ defined as 
\begin{align}
\delta\gamma_{N_{\alpha}}(\omega)=\frac{\left|\left(\textrm{Re}\,\gamma^{\uparrow\downarrow\uparrow\downarrow}_{\omega\omega';\nu}\right)_{N\alpha}-
\left(\textrm{Re}\,\gamma^{\uparrow\downarrow\uparrow\downarrow}_{\omega\omega';\nu}\right)_{N\alpha=10}\right|}{\left|\left(\textrm{Re}\,\gamma^{\uparrow\downarrow\uparrow\downarrow}_{\omega\omega';\nu}\right)_{N\alpha=10}\right|}.\label{dgamma}
\end{align}
are indicated as logarithmic plots in upper panel. About three digits accuracy can be found for the result with $N_\alpha$=2. The accuracy is improved to five digits for $N_\alpha$=4 and six digits or more for $N_\alpha \ge 6$. As mentioned in Sec.~\ref{LanczosED}, it is essential to include one large energy point $\Omega_\alpha \gg E_{\rm max}-E_l$, i.e., $\Omega_{N_\alpha}=5.12W$ in our example. Indeed, as shown by the dashed line in upper panel, the accuracy of the result drastically deteriorates from its counterpart, i.e., from six digits to less than three digits, when $\Omega_6$ is omitted in the calculation with the $N_\alpha = 6$ reference points.
 
\section{Comparison of DOS inferred by the present and standard MEMs}
\begin{figure*}
\includegraphics[width=4cm]{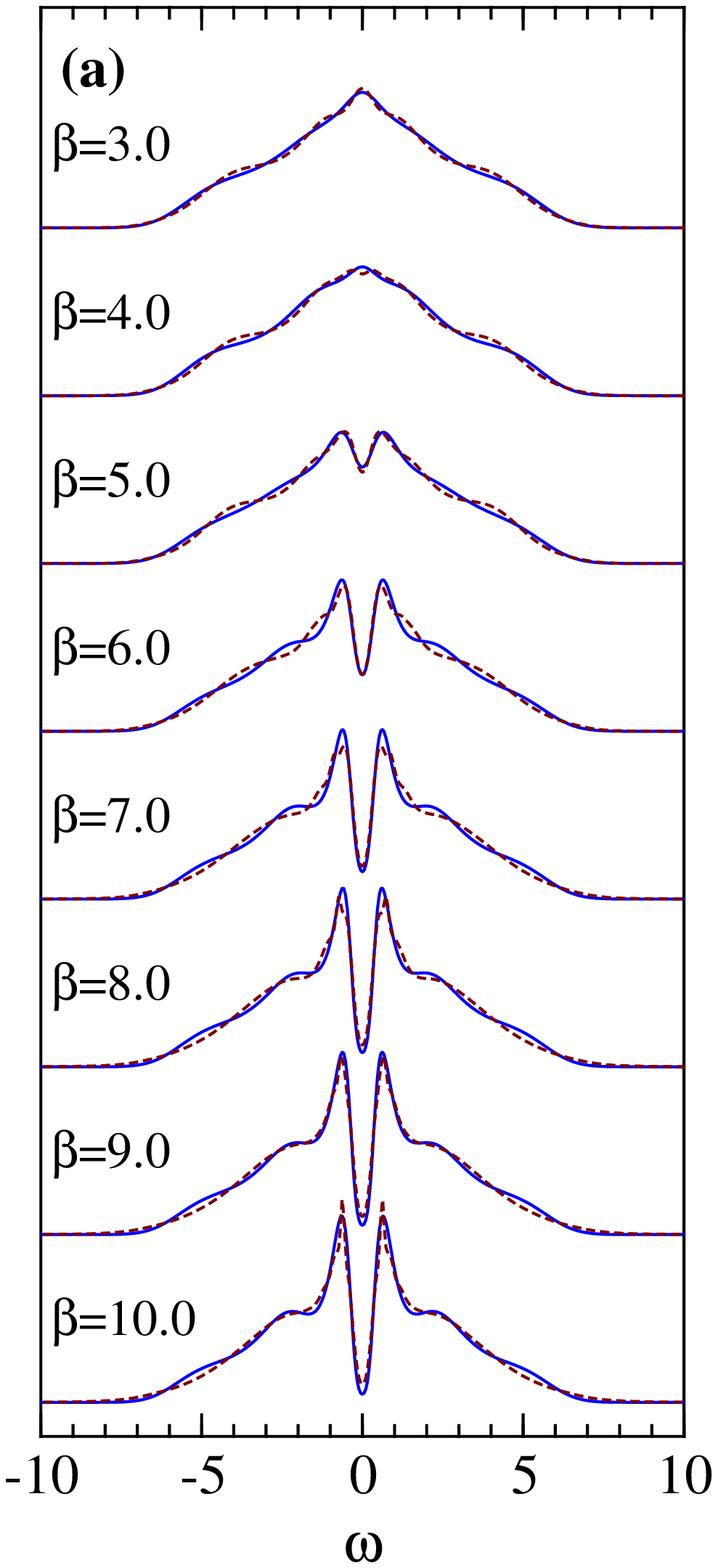}~
\includegraphics[width=4cm]{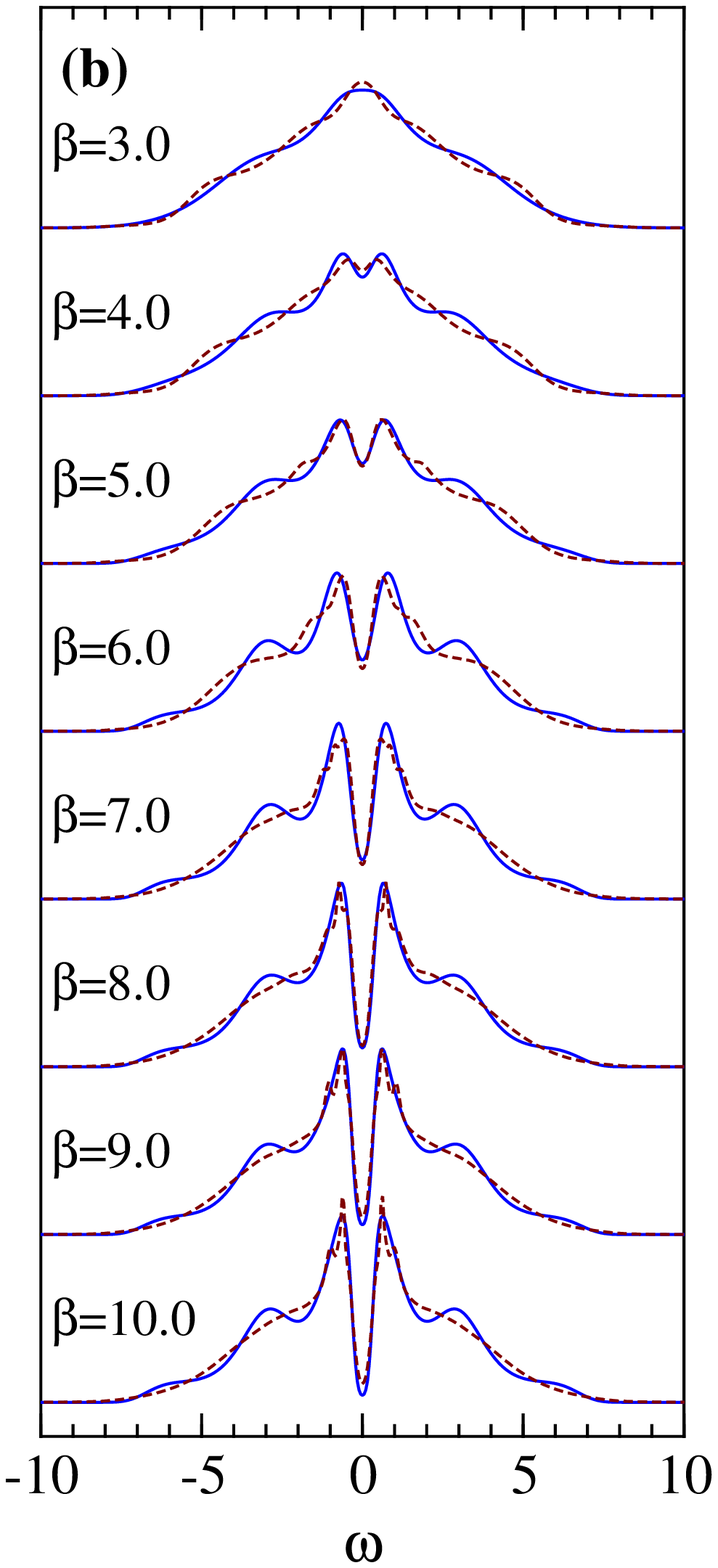}~
\includegraphics[width=4cm]{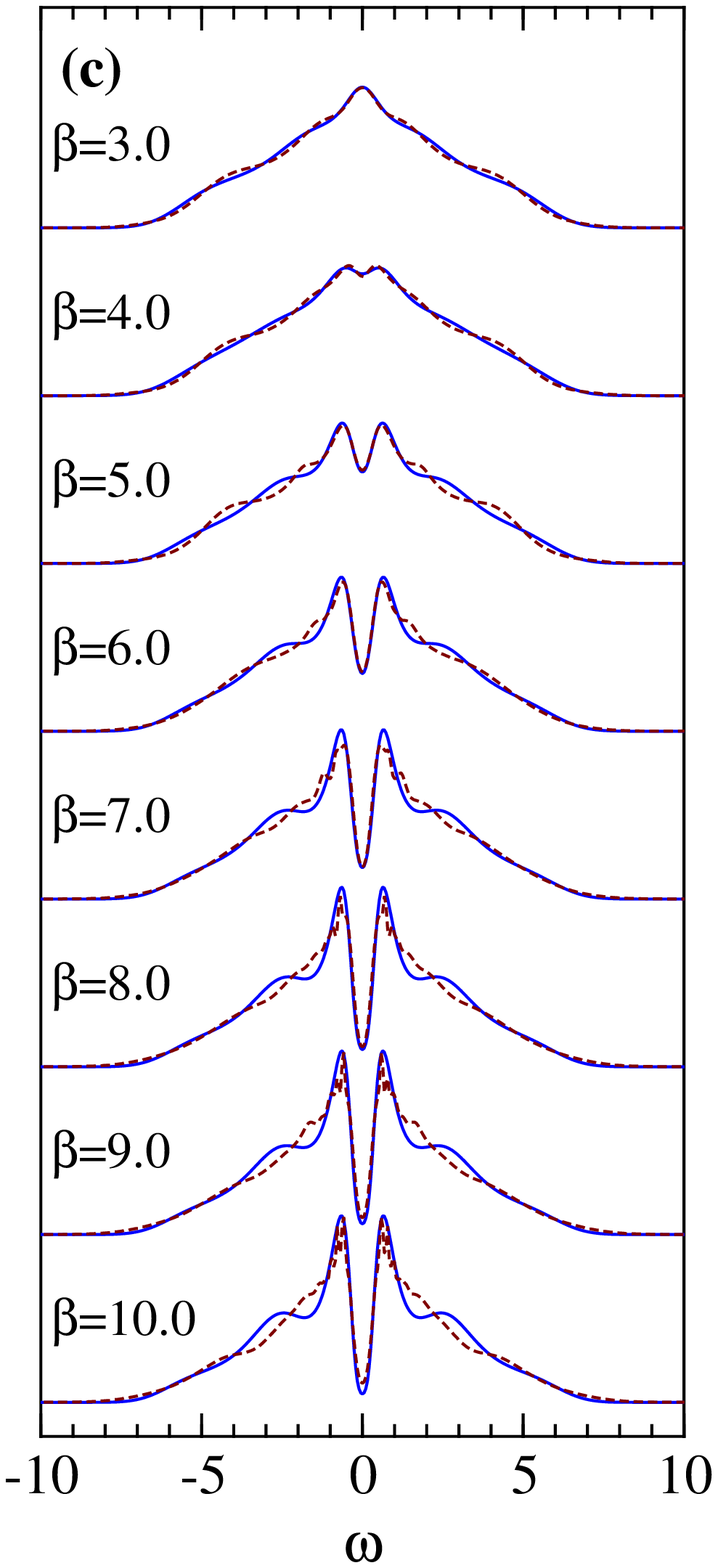}~
\includegraphics[width=4cm]{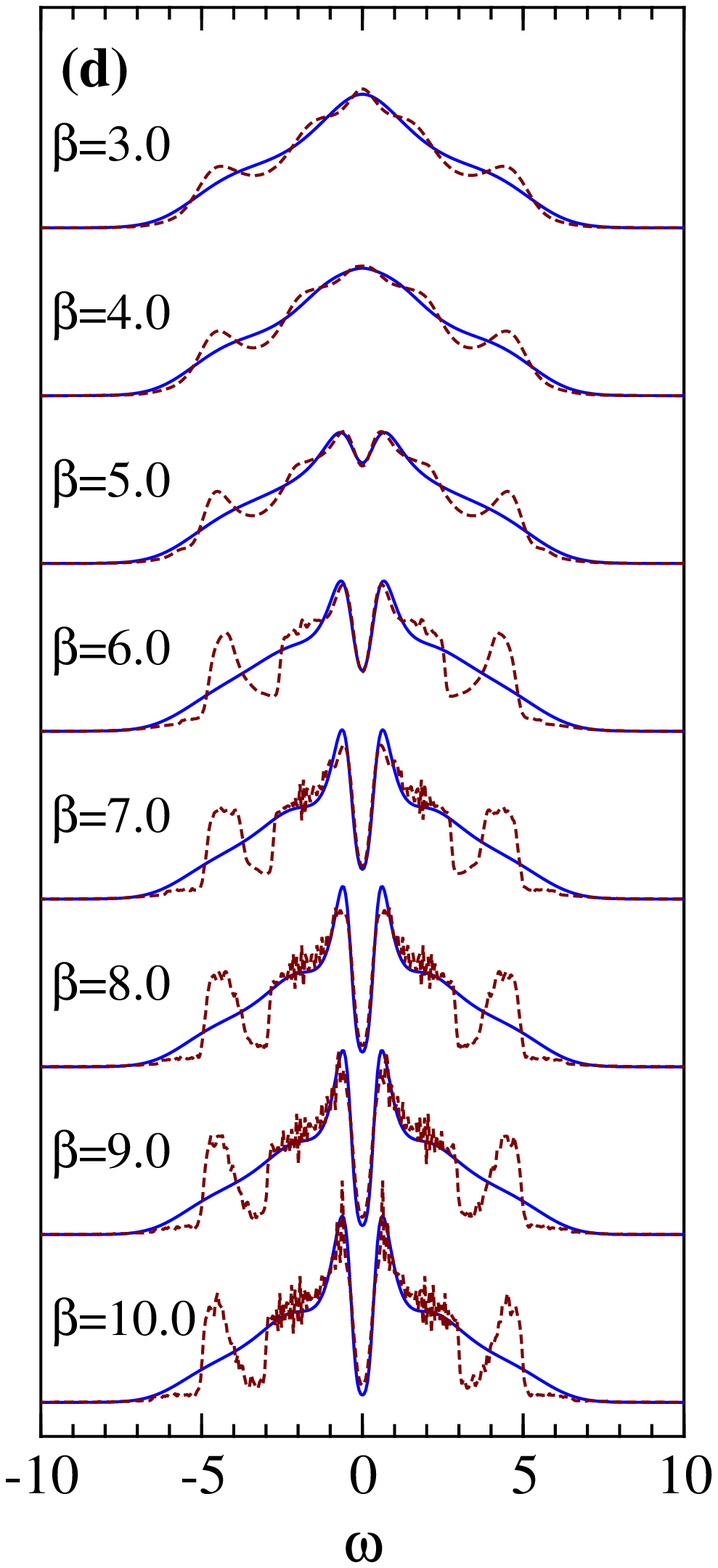}
\caption{\label{DOScmp}DOS $\rho(\omega)$ of the 2D Hubbard model for $U=4.0$ by means of LDFA with various value of $\beta$ obtained using the analytic continuation with the present MEM  (a) and with partly modified versions of the present MEM (b)-(d). In (b), only the data of the Green's function on the imaginary axis (the number of the data is $N_G=256$) are used instead of those indicated in Fig.~\ref{stadium}. In (c), the sum of squared absolute errors is assumed for $\chi^2$ instead of relative errors in Eq.\,(\ref{chi2}). In (d), the standard deviation in Eq.\,(\ref{PGA}) is fixed to $\sigma={\alpha_{\chi}}^{-1/2}=5\times 10^{-5}$ instead of optimizing hyperparameter $\alpha_{\chi}$ from the data. The solid lines are the DOSs directly inferred from $G_{\rm loc}(z_n)$ data by MEM and the dashed lines are those obtained from the summation of  $A_{\bm{k}}(\omega)$ inferred from corresponding $G_{\bm{k}}(z_n)$ data set by MEM.}
\end{figure*}
In the present study, the MEM adopted for the analytic continuation of the DOS and spectral functions is different from the standard method used in the majority of the previous studies\cite{MJarrell1996}. As mentioned in Sec.~\ref{MEM}, the differences can be boiled down to the three points: (i) not only the data of the Green's function on the imaginary axis but those on the complex plane indicated in Fig.~\ref{stadium} are used, (ii) the sum of squared relative errors in Eq.\,(\ref{chi2}) is adopted for $\chi^2$ instead of the sum of squared absolute errors, and (iii)  the standard deviation $\sigma={\alpha_{\chi}}^{-1/2}$ of the Gaussian function in Eq.\,(\ref{PGA}) is not given as a parameter but inferred from the data as a hyperparameter $\alpha_{\chi}$.

The purpose of this section is to clarify to what extent these differences affect the results.
The quality of the results can be assessed by comparing the DOS $\rho(\omega)$ directly inferred from the local Green's function $G_{\rm loc}(z_n)$ data by MEM to the DOS obtained from the summation of the spectral function $A_{\bm{k}}(\omega)$ inferred from corresponding $G_{\bm{k}}(z_n)$ data by MEM: $\rho(\omega)=(1/N)\sum_{\bm{k}}A_{\bm{k}}(\omega)$. $\rho(\omega)$ and $A_{\bm{k}}(\omega)$ are treated as $N_A=512$ discretized data within the range of $-1.5W <\omega <1.5W$. 

The LDFA results of DOS of the 2D Hubbard model with $U=4.0$ and various values of $\beta$ are shown in Fig.~\ref{DOScmp}.
The DOSs obtained with the present MEM are indicated in panel (a) and the results obtained with the points (i), (ii) and (iii) mentioned above being altered to those of the standard method are shown in panels (b), (c), and (d), respectively. The solid lines are the DOSs directly inferred from the $G_{\rm loc}(z_n)$ data by MEM and the dashed lines are those obtained from the summation of  $A_{\bm{k}}(\omega)$ inferred from corresponding $G_{\bm{k}}(z_n)$ data by MEM.

The DOSs obtained directly from the $G_{\rm loc}(z_n)$ data by the present MEM are in good agreement with those obtained from the summation of $A_{\bm{k}}(\omega)$ throughout the whole energy and temperature ranges as can be seen in panel (a). On the other hand, the deviations are apparent in the high-energy structures ($|\omega|>1$) of the results in panel (b) and also those at the low temperatures in panel (c). The deviations are much larger in panel (d) and this is because the quasiparticle-like peaks appeared at the high-energy region in the spectral function, e.g., those placed near the $\Gamma$ point in Fig.~\ref{AK},  get too sharp with fixed $\sigma=5\times 10^{-5}$ (estimated $\sigma$ at the $\Gamma$ point is one order larger in the case of the present MEM). 

As discussed in Sec.~\ref{LanczosED} and also demonstrated in Appendix C, the Lanczos ED method used in this study is the method which is accurate not only in the low energy properties but also in the asymptotic behavior. Because of this feature, it is expected that one can infer more accurate DOS or spectral functions by exploiting high-energy information of the data: by using relative errors instead of absolute errors in Eq.\,(\ref{chi2}) to give more weight to the data points on the high-energy side or by placing the data points not too far from the poles on the real axis as in Fig.~\ref{stadium} to avoid the loss of information. This explains why the present MEM can extract more accurate information of the DOS and spectral functions than the standard MEM.


\end{document}